\documentclass[aps,twocolumn,superscriptaddress]{revtex4}

\usepackage{amsmath,amssymb,bm}
\usepackage{graphicx}
\usepackage{amsfonts}
\usepackage{color}
\usepackage{color}

\newcommand{\nn}{\nonumber}                                           
\newcommand{\va}[1]{\langle{#1}\rangle}                               

\newcommand{\al}{\alpha}
\newcommand{\be}{\beta}
\newcommand{\ga}{\gamma}
\newcommand{\de}{\delta}

\newcommand{\si}{\sigma}

\newcommand{\Tr}{\mathop{\rm Tr}\nolimits}

\begin{document}

\thispagestyle{empty}
\date{\today}
\preprint{\hbox{...}}

\title{Hidden-charm pentaquarks with color-octet substructure
    in QCD Sum Rules}

 \author{Alexandr Pimikov}
\email{pimikov@mail.ru}
\affiliation{Institute of Modern
    Physics, Chinese Academy of Science, Lanzhou 730000, China}
\affiliation{Research Institute of Physics, Southern Federal University, Rostov-na-Donu 344090, Russia}

\author{Hee-Jung Lee}
\affiliation{Department of Physics Education, Chungbuk National
    University, Cheongju, Chungbuk 28644, Korea}

\author{Pengming Zhang}
\affiliation{School of Physics and Astronomy, Sun Yat-sen University, Zhuhai 519082, China}
\affiliation{Institute of Modern Physics,
    Chinese Academy of Science, Lanzhou 730000, China}

\begin{abstract}
We study the hidden-charm pentaquark states $udsc\bar{c}$ with
spins 1/2, 3/2, and 5/2 within the QCD sum-rule approach. 
First, we  construct the currents for the particular configuration of
pentaquark states that consist of the flavor singlet three-quark
cluster $uds$ of spins 1/2 and 3/2 and the two-quark cluster $\bar
cc$ of spin 1, where both clusters are in a color-octet state.
From the QCD sum rules obtained by the operator product expansion
up to dimension-10 condensates,
    the extracted masses for the pentaquark states $uds$-$\bar{c}c$
are about 4.6~GeV (5.6~GeV) for spin 1/2$^\pm$, about 5.1~GeV
(6.0~GeV) for spin 3/2$^\pm$, about 6.1~GeV (5.9~GeV) for spin
5/2$^\pm$, where the masses of the positive parity states are
given in parentheses. Additionally, based on the flavor singlet
pentaquark states, it is also shown that other pentaquark states
of clusters like $udc$-$\bar{c}s$ and $usc$-$\bar cd$ lead to
masses similar to the $uds$-$\bar cc$ case within error bars.
Furthermore, in order to see whether any of the states, observed
by the LHCb Collaboration, could be understood as the pentaquark
of two clusters in the color-octet state, we study the pentaquark
formed by the two clusters $udc$-$\bar cu$, where the three-quark
cluster is assumed to have the same flavor structure as the above
$uds$ cluster. We come to the conclusion that if the observed
pentaquark will be found to have spin 1/2 and negative parity,
then it could be described as a state of two color-octet clusters.
\end{abstract}

\maketitle

\section{Introduction}

Since the observation of the two exotic hidden-charm pentaquark $P_c^+$ states of
the quark content $uudc\bar{c}$ with the spins 3/2 and 5/2 through the decay
$\Lambda_b^0\rightarrow J/\psi K^- p$ by LHCb collaboration~\cite{Aaij:2015tga},
many studies on these states and other expected hidden-charm pentaquark states have been performed.
Note that recently the LHCb collaboration has observed a
three peak structure~\cite{Aaij:2019vzc,Cao:2019kst} using an updated analysis.
There is an intriguing possibility~\cite{Cao:2019gqo,Wang:2019krd}
that pentaquark states could be observed
on Electron-Ion Collider China (EicC). 
The pentaquark states, including the above hidden-charm pentaquark states, 
have been theoretically studied using
quark models~\cite{Wu:2017weo, Santopinto:2016pkp, Irie:2017qai},
diquark models and triquark-diquark models~\cite{Maiani:2015vwa, Lebed:2015tna, Li:2015gta, Zhu:2015bba, Wang:2015epa,Jaffe:2003sg,Sugiyama:2003zk,Lee:2005ny,Karliner:2003dt,Zhu:2003ba},
hadronic molecular states~\cite{Chen:2015moa, Roca:2015dva, He:2015cea, Azizi:2016dhy, Guo:2017jvc},
the coupled-channel unitary approach~\cite{Wu:2010jy, Wu:2010vk, Shen:2019evi},
the contact-range effective field theory~\cite{Liu:2019tjn},
and the hadroquarkonia model \cite{Eides:2017xnt}. 
For a review on the hidden-charm multiquark states, see~\cite{Chen:2016qju}.
Among the expected hidden-charm pentaquark states, it is very intriguing to
analyze the pentaquark state of the quark content $udsc\bar{c}$
since $\Lambda_b^0$ could also decay into $J/\psi K^- p$ via $J/\psi \Lambda^*$ and
such pentaquark state could be observed through the decay $\Xi^-_b\rightarrow J/\psi\Lambda K^-$~\cite{Santopinto:2016pkp}.

Most of the models and approaches to the pentaquark states in the references 
discussed above rely on an assumption
that a pentaquark state under consideration has
certain structure in color, spin, and flavor.
Considering the interpolating current to a pentaquark state for analysis within the QCD sum-rules (SRs),
the clustering in the color, flavor and spin space is inevitable due to the
absence of the invariant rank-5 tensors for the color, flavor and spin subspaces.
For example for $SU(3)_c$, the largest rank of invariant tensor is 3, therefore considering 
two-quarks and three-quarks cluster
 would be one of the natural possibilities in constructing the interpolating for a pentaquark state.
Recently, the hidden-charm pentaquark state of the quark content $udsc\bar{c}$ with the flavor singlet structure in $SU_F(3)$
(the flavor singlet hidden-charm pentaquark)
was considered in~\cite{Irie:2017qai} within quarks models.
Specially, the flavor singlet hidden-charm pentaquark was analyzed as the bound state of a three-quark
and two-quark parts both in color octets, and the stable result was got for the total spin 1/2 in~\cite{Irie:2017qai}.

In this paper,
we study first the flavor singlet hidden-charm pentaquark states
of $udsc{\bar c}$ with the spins $1/2$, $3/2$, and $5/2$
using QCD SRs.
We assume that these states consist of two colored clusters as
discussed in~\cite{Mironov:2015ica,Irie:2017qai,Takeuchi:2016ejt} within quark models.
So, we consider them as states consisting of
the three-quark cluster $uds$ and the two-quark cluster ${\bar c}c$.
Additionally, we assume that all quarks are in an $S$-wave, the colors of both clusters are color octets,
and the two-quark cluster has spin 1 since it has been shown in~\cite{Irie:2017qai} that such clusters
of $uds$ and ${\bar c}c$ yielded the most stable result.
To check this assumption, we consider the pentaquark states containing a scalar two-quark cluster
and find that such states lead to higher masses than those obtained from pentaquarks with a two-quark cluster of spin 1.
Then, we also examine other possible pentaquark states containing the two clusters
$udc$-${\bar c}s$ and $usc$-${\bar c}d$
by extending the results of the flavor singlet $uds$ and ${\bar c}c$ case.
Furthermore, the pentaquark states of the two color-octet clusters $udc$-${\bar c}u$ are studied to see if
any of the states observed by LHCb could be understood
in terms of pentaquark state with color-octet substructure
and flavor-singlet flavor three-quark part $udc$.
Let us point out that the method of QCD SR relies on the local current
	for studying the spectroscopy of hadrons.
In this work, therefore, we construct only the local currents for pentaquark in the configuration space
	with all quarks located at the same point.
	The particular configuration and clustering reflect
	the properties only in flavor, color and spin subspaces.

This paper is organized as follows.
Using the above assumptions, we construct in Sec. II
the interpolating currents for the hidden-charm pentaquark states
with spins 1/2, 3/2, and 5/2 in the form of a product of the currents for these two clusters as
\begin{eqnarray}\nn
J_\text{5q}=J_{3q}^m \, J_{2q}^m\,
\end{eqnarray}
with the color index $m$.
We perform the operator product expansion (OPE) for the correlators
with the interpolating currents in Sec. III and present the system of the
employed QCD sum rules in Sec. IV.
Furthermore, since the relativistic interpolating currents for the fermions can be coupled to the two states
with opposite parities when the QCD sum rules are constructed, we discuss how to extract the contribution
to a state with definite parity from the system of the QCD sum rules in Sec. IV.
Finally, a comprehensive discussion of the results is given in Sec. V.

\section{Interpolating Currents}
\label{sec:interpolating.J}

First, we consider the wave function of the flavor singlet $uds$ cluster for constructing
the three-quark interpolating current $J_{3q}^m$.
Then, we extend our current to the case of three arbitrary flavors.
We will take the flavor structure of the interpolating current $J_{3q}^m$ from the flavor singlet wave function in the flavor $SU(3)$ space as
\begin{eqnarray}\label{eq:flavorCurrent}
J_\text{flavor}\sim(ud-du)s+(su-us)d+(ds-sd)u\, .
\end{eqnarray}
We study both cases of an $uds$ cluster: one with spin 1/2 and one with
spin 3/2 for a total spin 1/2, 3/2, and 5/2 of the pentaquark states.
To this end, we adopt the QCD SR method applied to the analysis of the baryon octet \cite{Ioffe:1981kw}.
Therefore, we construct the current with the first two quarks contributing spin 0
to the total spin of the three-quark cluster with spin 1/2.
On the other hand, in the current for the three-quark cluster with spin 3/2, 
the first two quarks give spin 1 to the total spin.

With these ingredients and Ioffe's current~\cite{Ioffe:1981kw,Ioffe:2010zz} with 
the definite chiralities which are well known to form a good basis,
we consider the following structure of the spin part of the interpolating current of
the three-quark cluster with spin~1/2
\begin{eqnarray}\label{eq:JA}
J_\text{spin}^\text{A}\!&=&\! 4(u^{T}_{R} C \Gamma_\text{A} d_{R})\Gamma_2 s_{L}-
(R\leftrightarrow L)\\\nn &=&
(u^TC\{\gamma_5,\Gamma_\text{A}\}d) \Gamma_2s- (u^TC\{\gamma_5,\Gamma_\text{A}\}\gamma_5d)\Gamma_2 \gamma_5 s\, ,
\end{eqnarray}
where the superscript A means that the current is antisymmetric under the exchange
of the spinor indexes of the first two quarks.
The first term in Eq.~(\ref{eq:flavorCurrent}) is considered as an example, and 
then the rest is included in the final stage.
From the above expression, $\Gamma_\text{A}$ must 
satisfy the following conditions in order to have no zero current
\begin{equation}\nn
\{\gamma_5,\Gamma_\text{A}\}\neq 0\ ,\ (C\Gamma_\text{A})^T=-C\Gamma_\text{A}\,,
\end{equation}
where $T$ means the transposition.
These conditions limit the choices of $\Gamma_\text{A}$ to $\Gamma_\text{A}=1,\ \gamma_5$.
For a $uds$ cluster of spin~3/2, we consider
\begin{eqnarray}\label{eq:JS}
J_\text{spin}^\text{S}\!&=&\! 2(u^{T}_{R} C \Gamma_\text{S} d_{L})\Gamma_2\ga_5 s+(R\leftrightarrow L)\\\nn &=&
(u^TC[\Gamma_\text{S},\gamma_5]\ga_5 d) \Gamma_2 \ga_5 s\,
\end{eqnarray}
where the superscript S denotes that the current is symmetric under the exchange
of the spinor indexes of the first two quarks.
Similarly to the case of the above current for the spin 1/2 case,
$\Gamma_\text{S}$ must satisfy the following conditions
\begin{equation}\nn
[\gamma_5, \Gamma_\text{S}]\neq 0, \ (C\Gamma_\text{S})^T=C\Gamma_\text{S}\,.
\end{equation}
in order to have a nonzero current.
Therefore, the only choice is $\Gamma_\text{S}=\gamma_\mu$.

Before constructing the full current, we study the currents in color subspace.
Using the adjoint representation of color $SU(3)$, the color-octet structure of the  current can be constructed as
\begin{eqnarray}\nn
J_\text{color}^\text{S/A} = \epsilon_{ac\omega}t^m_{\omega b} ~u_{a}d_{b}s_{c} \,,
\end{eqnarray}
where $m$ is a color index.
Other choices for color tensors lead to zero currents or
to the same full currents due to the symmetries in the spin and flavor subspaces.

To generalize the $uds$ case, considered above, to other flavors of three-quark clusters, we
distinguish the quarks in the three-quark cluster by $(q_1,q_2,q_3)$.
Combining the currents constructed in the flavor, spin and color subspaces,
we get the interpolating currents for the considered structure of pentaquark states
in the form of
\begin{eqnarray}\label{eq:Jfull}
J_{l}^\text{A}(\Gamma_2\,,\Gamma_3)=T_{l,m}^\text{A}(\Gamma_2)
q^T_1 q_2 q_3 (\bar q_5t^m\Gamma_3 q_4)\,,\\\nn
J_{\mu,l}^\text{S}(\Gamma_2\,,\Gamma_3)=T_{\mu,l,m}^\text{S}(\Gamma_2)
q^T_1 q_2 q_3 (\bar q_5t^m\Gamma_3 q_4)\,.
\end{eqnarray}
Here, the quark fields in the three-quark cluster carry flavor $f_i$, color $c_i$ and spin $l_i$ indices
as $q_i=q_{f_ic_il_i}$
to be contracted with the tensor which becomes
\begin{eqnarray} \nn
T^\text{A}_{l,m}(\Gamma_2)&=&
\large(
C_{l_1l_2}(\Gamma_2)_{ll_3}-(C \gamma_5)_{l_1l_2}(\gamma_5 \Gamma_2)_{ll_3}
\large)\\\nn
&&\times
\epsilon_{f_1f_2f_3}\epsilon_{c_1c_3c}t^m_{c c_2}\,,
\\\nn
T^\text{S}_{\mu,l,m}(\Gamma_2)&=&
\large(C\ga_\mu\large)_{l_1l_2}
\large(\Gamma_2\large)_{ll_3}
\epsilon_{f_1f_2f_3}\epsilon_{c_1c_3c}t^m_{c c_2}\,.
\end{eqnarray}
These definitions reflect our choice for $\Gamma_\text{A}=\gamma_5$ and $\Gamma_\text{S}=\gamma_\mu$.
We denote the flavor configuration by
$q^1q^2q^3\text{-}\bar q^5q^4$,
where $q^i$ is the flavor of the $i$-th quark $q_i$.
In this work, as mentioned in the introduction, we consider four cases for a given flavor configuration:
$uds$-$\bar cc$\,,~
$udc$-$\bar cs$\,,~
$usc$-$\bar cd$ and
$udc$-$\bar cu$.
Quarks fields are contracted with the antisymmetric tensor of flavor indices that
corresponds to the flavor singlet configuration.
Note that the free index $l$ denotes the spinor component of the current and will be omitted in the following discussion.

We mention here again that the spinor structure of the three-quark cluster in the full current Eq.~(\ref{eq:Jfull})
is chosen to have the particular structure of
$q_Rq_Rq_L-q_Lq_Lq_R$ for the antisymmetric case, Eq.~(\ref{eq:JA}), and
$q_Rq_Lq+q_Lq_Rq$ for the symmetric case Eq.~(\ref{eq:JS}).

The matrix $\Gamma_2$, which can be considered as a factor of the current due to the following properties
\begin{eqnarray}\nn
J^\text{S/A}_l(\Gamma_2\,,\Gamma_3)=(\Gamma_2)_{lk}J^\text{S/A}_k(1,\Gamma_3)\, ,
\end{eqnarray}
will be chosen according to the P-parity and the spin of the interpolating  current under consideration.
As for $\Gamma_3$, following the analysis of~\cite{Irie:2017qai}, where
the two-quark cluster with spin 1 in the $uds$-${\bar c}c$ system yielded the most stable result,
we will take $\Gamma_3=\ga_\nu$ for most pentaquark states considered in this work.
An alternative option for $\Gamma_3=1$ will also be considered.

To discuss the symmetry properties of the constructed three-quark currents in the color-spin subspace,
we consider the six-dimensional fundamental representation of the $SU(6)$ group ~\cite{Gursey:1992dc,bookCloseQuarks} composed of the tensor product
of the color $SU(3)_{\rm color}$ and the spin $SU(2)_{\rm spin}$ subgroups.
Representing a quark by its dimension $(3,2)$ where the first (second) corresponds to the dimension of
$SU(3)_{\rm color}$ ($SU(2)_{\rm spin}$) subgroup, we have
\begin{eqnarray}\nn
\!\!(3,2)\!\otimes\!(3,2)\!\otimes\!(3,2)=
(8,2)\!\oplus\!(8,4)\!\oplus\cdots\,,
\end{eqnarray}
where only two irreducible representations are shown.
The first term on the
right-hand side has spin 1/2 and belongs to the fully symmetric 56-plet representation,
while the second term has spin 3/2 and belongs to the mixed symmetric 76-plet of the full $SU(6)$ group.
The color-spin part of the constructed current $J^\text{A}$, Eq.~(\ref{eq:Jfull}), represents
(8,2) states studied in~\cite{Irie:2017qai}.
The current $J^\text{S}$ corresponds to (8,2) and (8,4), depending on the choice of $\Gamma_2$.

In this section, we have constructed the general form for the pentaquark currents $J_8\sim qqq$-$\bar qq$ with two color-octet compounds.
The suggested currents are unique and can't be presented by the sum of any other currents considered previously.
Nevertheless, omitting the flavor structure, we have related this type of current with 
another types of currents that represent following configurations in color subspace:
    diquark-diquark-antiquark clustering $J_{\bar 3}\sim qq$-$qq$-$\bar q$
    with an anti-triplet color substructure suggested in \cite{Jaffe:2003sg,Sugiyama:2003zk,Lee:2005ny},
    and a molecule form $J_1\sim qqq$-$\bar qq$ with color-singlet parts, see~\cite{Guo:2017jvc}.
We conclude, that for any current of color-octet type $J_8$,
one can find two specific (in spin and isospin) currents of color-singlet type
$J_1$ and color-anti-triplet type $J_{\bar 3}$ such that $J_8=J_1+ J_{\bar 3}$.
For more details see Appendix C, where we show how to construct these specific currents in 
spin space.

\section{OPE for $1/2$, $3/2$, $5/2$-states}
\label{sec:OPE}

The correlator  $\Pi^s_{(\mu)(\nu)}(q^2)$ for the QCD sum-rule analysis of a pentaquark state is defined by
\begin{equation}\label{eq:correlator}
\Pi^s_{(\mu)(\nu)}(q^2)=i\int d^4x e^{iq\cdot x}\langle0|TJ_{(\mu)}(x)\bar{J}_{(\nu)}(0)|0\rangle
\end{equation}
with the interpolating current $J_{(\mu)}$ for the considered pentaquark state of spin $s$.
The subscript $(\mu)$ stands for the possible Lorentz indices of currents
for the $s=3/2,~ 5/2$ states.
Since the current $J_{(\mu)}$ can couple to the states with a spin lower than $s$,
the phenomenological part of the SRs
contains contributions from the lower spin states as well.
Extracting the contribution from the state with spin $s$ only, the correlator
can be written as
\begin{equation} \label{eq:Pi-Sterm}
\Pi^s_{(\mu)(\nu)}(q^2)=(\hat{q}\Pi^s_1(q^2)+\Pi^s_2(q^2))S^s_{(\mu)(\nu)}+\cdots\,,
\end{equation}
where $\hat{q}=\gamma\cdot q$ and $\cdots$ means the terms corresponding
to the omitted contributions from states with spin $s$ and also lower spins.
Therefore, to construct SRs for the state of spin $s$, one needs to extract $\Pi^s_{1,2}$ from the correlator.
The ways of extracting $\Pi^s_{1,2}$ for $s=3/2,~5/2$ are summarized in Appendices A and B.
Then, QCD SRs for the state of the spin $s$ will be constructed by applying the dispersion relation~\cite{Shifman:1978bx}
to the two scalar functions $\Pi^s_{1,2}$ in Eq.~(\ref{eq:Pi-Sterm})
\begin{eqnarray}\label{eq:SR-disp}
\Pi^s_i(q^2)=\int_{s_\text{th}}^{\infty}dt \frac{\rho^s_i(t)}{t-q^2}\,.
\end{eqnarray}
Here the spectral densities $\rho^s_i(t)$ are defined in the physical $t$ region by
\begin{eqnarray}\label{eq:density}
\rho^s_i(t)=\frac{1}{\pi}\text{Im}\Pi^s_i(t)
\end{eqnarray}
with $i=1,2$.

In the next subsections A, B, C, we present the relativistic interpolating currents for each state of spin 1/2, 3/2, and
5/2 states with proper choices of $\Gamma_1$, $\Gamma_2$, and $\Gamma_3$.
Then, in subsection D, we show how to calculate the
spectral densities $\rho^s_i(t)$ within the OPE for the QCD sum rules for each state.

\subsection{$J^P=1/2^{\pm}$-states}

We consider four types of the current for the spin 1/2 case:
\begin{eqnarray}\label{eq:current12types}
&&
J^1=J^\text{A}(\gamma_5\gamma_\mu\,,\gamma_\mu)\,,~~
J^2=J^\text{S}_\mu(\gamma_5\,,\gamma_\mu)\,,
\\\nn&&
J^3=J^\text{A}(\gamma_5\,,1)\,,~~
J^4=J^\text{S}_\mu(\gamma_5\gamma_\mu\,,1)\,,
\end{eqnarray}
where the upper index denotes the type of the current.
The main results for the spin 1/2 case are obtained using the current $J^1$,
while currents $J^2$, $J^3$, $J^4$ are also studied as an alternative
option.
The interpolating current $J^1$ with the quantum numbers $1/2^{+}$ can be related
to the spin-3/2 current as follows
\begin{eqnarray}\nn
J^1=J^\text{A}(\gamma_5\gamma_\mu\,,\gamma_\mu)=-\gamma_\mu J^1_\mu\, .
\end{eqnarray}
The choice of $\Gamma_2=\gamma_5\gamma_\mu$ insures that the spin-3/2 current $J_\mu$
is projected by $\Gamma_2$ only on the 1/2-spin component so that $\va{0|\gamma_\mu J_\mu|3/2^\pm}\sim \ga_\mu u_\mu =0$
thanks to the subsidiary condition for the 3/2 spinor $u_\mu$
(see Eqs. (\ref{eq:current32types}) and (\ref{eq:spinSum32})).
Since the relativistic interpolating current is considered, as discussed in~\cite{Chung:1981cc,Jido:1996ia},
the current can couple to the state of negative parity as well.
Denoting two such states by $|1/2^+\rangle$ and $|1/2^-\rangle$,
the current couples to the states through the following relations
\begin{eqnarray}\nn
\va{0|J|1/2^+} &=& f_{\frac 12+} u\,,~~~
\va{0|J|1/2^-} = f_{\frac 12-}\gamma_5 u\,,\\\label{eq:spinSum12}
\sum\limits_{s}u(q,s)\bar u(q,s)&=&\hat q+m\
\end{eqnarray}
with the spinor $u$. The structure of the correlator becomes
\begin{eqnarray}\nn
\Pi^{1/2}(q^2)&=&\hat{q}\Pi^{1/2}_1(q^2)+\Pi^{1/2}_2(q^2)
\end{eqnarray}
and then $S^{1/2}_{(\mu)(\nu)}=1$ because there is no Lorentz index in the current.
The two spectral densities can be obtained as
\begin{eqnarray}\nn
\rho_1^{1/2}(s)&=&\frac{1}{4\pi s}{\rm Tr}\bigg(\hat{q}\textbf{Im}\Pi^{1/2}(s)\bigg)\ ,
\\\nn
\rho_2^{1/2}(s)&=&\frac{1}{4\pi}{\rm Tr}\bigg(\textbf{Im}\Pi^{1/2}(s)\bigg)\ .
\end{eqnarray}

\subsection{$3/2^{\pm}$-states}

For spin 3/2 states , we study two types of the current
    \begin{eqnarray}\label{eq:current32types}
    &&J^1_\mu=J^\text{A}(\gamma_5\,,\gamma_\mu)\,,~~J^2_\mu=J^\text{S}_\mu(\gamma_5\,,1)\,.
    \end{eqnarray}
The main results will be obtained by using the current $J^1_\mu$ that
has the quantum numbers $3/2^{-}$.
As in the spin-1/2 case, the interpolating current couples to the states of both parities
through the relations with the corresponding spinors $u_\mu$~\cite{Azizi:2016dhy,Wang:2015epa}
\begin{eqnarray}\nn
&&\va{0|J_\mu|3/2^+} = f_{\frac 32+}\gamma_5 u_\mu\,,~~~
\va{0|J_\mu|3/2^-} = f_{\frac 32-} u_\mu\,,\\\label{eq:spinSum32}
&&\sum\limits_{s}u_\mu(q,s)\bar u_\nu(q,s)=(\hat q+m)T_{\mu\nu}\,,
\end{eqnarray}
where the tensor $T_{\mu\nu}$ is
\begin{eqnarray}\nn
T_{\mu\nu}&=& -g_{\mu\nu}+\frac 13 \gamma_\mu\gamma_\nu+\frac{2q_\mu q_\nu}{3q^2}-
\frac{q_\mu\gamma_\nu-q_\nu\gamma_\mu}{3\sqrt{q^2}}\ .
\end{eqnarray}
Note that $\gamma_5$ in the first relation in Eq.~(\ref{eq:spinSum32}) appears because the current has an intrinsic negative parity.
The correlator has the structure
\begin{eqnarray}\nn
\Pi^{3/2}_{\mu\nu}(q^2)&=&
\bigg(\hat{q}\Pi^{3/2}_1(q^2)+\Pi^{3/2}_2(q^2)\bigg)(-g_{\mu\nu})
+\cdots\ .
\end{eqnarray}
Since it is known that the pure contributions from the $S=3/2$ state to the correlator
can be defined  by the terms proportional to $S^{3/2}_{(\mu)(\nu)}=-g_{\mu\nu}$
~\cite{Wang:2015epa,Leinweber:1989hh,Azizi:2016dhy},
we show only the relevant terms here.
The other terms that contribute to the correlator are given in Appendix~\ref{app:32projectors}
together with the derivation of the exact form for the projectors $P^{3/2,i}_{\mu\nu}$.
As in Appendix A, the two spectral densities can be obtained as
\begin{eqnarray}\label{eq:rho-32}
\rho^{3/2}_1(s) &=& \frac 1\pi {\rm Tr}\bigg[\textbf{Im}\Pi^{3/2}_{\mu\nu}(s) P^{3/2,1}_{\mu\nu}\bigg]\,,
\\\nn
\rho^{3/2}_2(s) &=& -\frac 1\pi {\rm Tr}\bigg[\textbf{Im}\Pi^{3/2}_{\mu\nu}(s) P^{3/2,2}_{\mu\nu}\bigg]\,.
\end{eqnarray}
More explicit forms are presented in Eq.~(\ref{eq:rho-32-mod}).
Here, we point out that an extra factor -1 is introduced in $\rho_2^{3/2}$ for the construction
of the SRs in one single form for all spin cases. 
This factor is related to the intrinsic negative parity of the current, see Eq.~(\ref{eq:spinSum32}).

\subsection{$5/2^{\pm}$-states}

The only type of current studied here is
\begin{eqnarray}\label{eq:current52types}
J^1_{\mu\nu}=J^\text{S}_\mu(\gamma_5\,,\gamma_\nu)+(\mu \leftrightarrow\nu)\,,
\end{eqnarray}
with the choice $\Gamma_2=\gamma_5$ corresponding to the quantum numbers $5/2^{+}$.
This current couples to the states of both parities through the relations
~\cite{Azizi:2016dhy,Wang:2015epa}:
\begin{eqnarray}\nn
&&\va{0|J_{\mu\nu}|5/2^+} = f_{\frac 52+}u_{\mu\nu}\,,~~
\va{0|J_{\mu\nu}|5/2^-} = f_{\frac 52-}\gamma_5 u_{\mu\nu}\,,\\\label{eq:spinSum52}
&&\sum\limits_{s}u_{\mu\nu}(q,s)\bar u_{\al\be}(q,s)=(\hat q+m)T_{\mu\nu,\al\be}\,,\\\nn
&&T_{\mu\nu,\al\be}\equiv
\frac{\tilde g_{\mu\al}\tilde g_{\nu\be}+\tilde g_{\mu\be}\tilde g_{\nu\al}}{2}
-\frac{\tilde g_{\mu\nu}\tilde g_{\al\be}}{5}
-\frac{2}{5}t_{\{\mu\nu\},\{\al\be\}}\,,\\\nn
&&t_{\mu\nu,\al\be}=
\left(
\gamma_\mu\gamma_\al
-\frac{q_\mu\gamma_\al-q_\al\gamma_\mu}{\sqrt{q^2}}
-\frac{q_\mu q_\al}{q^2}
\right)\tilde g_{\nu\be}\,,
\end{eqnarray}
where symmetrization of the two indices in the curly brackets in the tensor $t$
is imposed by $t_{\{\mu\nu\}}=t_{\mu\nu}+t_{\nu\mu}$. 
The corresponding correlator has a rather complicated structure as
one can see from~\cite{Wang:2015epa}.
We calculate those terms known
to contribute to the correlator only from the spin-5/2 state~\cite{Azizi:2016dhy,Wang:2015epa} as
\begin{eqnarray}\nn
\Pi^{5/2}_{\mu\nu,\al\be}(q^2)&=&
\bigg(\hat{q}\Pi_1^{5/2}+\Pi_2^{5/2}\bigg)\frac{(g_{\mu\alpha}g_{\nu\beta}+g_{\mu\beta}g_{\nu\alpha})}{2}+\cdots\,.
\end{eqnarray}
Therefore,
$S^{5/2}_{(\mu)(\nu)}=
(g_{\mu\al}g_{\nu\be}+g_{\mu\be}g_{\nu\al})/2$
and we calculate the two spectral densities ($i=1,2$) through
\begin{eqnarray}\label{eq:rho-52}
\rho^{5/2}_i(s)&=&\frac 1\pi \Tr(\textbf{Im}\Pi^{5/2}_{\mu\nu,\al\be}(q^2)P^{5/2,i}_{\mu\nu,\al\be})\,,
\end{eqnarray}
where the projectors $P^{5/2,i}_{\mu\nu,\al\be}$ are constructed in Appendix~\ref{app:52projectors},
see Eq.~(\ref{eq:52projector}).

\subsection{OPE of correlators}

In the previous subsections,
    we constructed various currents for spin 1/2, 3/2, and 5/2 pentaquark states.
We specify the current by its three properties:
    (i) the spin of the pentaquark (1/2, 3/2, 5/2),
    (ii) the flavor clustering ($uds$-$\bar cc$, $udc$-$\bar cs$, $usc$-$\bar cd$, $udc$-$\bar cu$) of the current,
    (iii) the type of the current.
    For spin-1/2, we have introduced four options (type-1,2,3,4), for spin 3/2 -- two (type-1,2), for spin 5/2 -- only one current type-1.
The following considerations of this subsection and
the next section are based on the general definition of the correlator,
Eq.~(\ref{eq:correlator}), and are
relevant to any current considered in the previous subsections.

\begin{figure}[t]
	\includegraphics[height=0.1\textwidth]{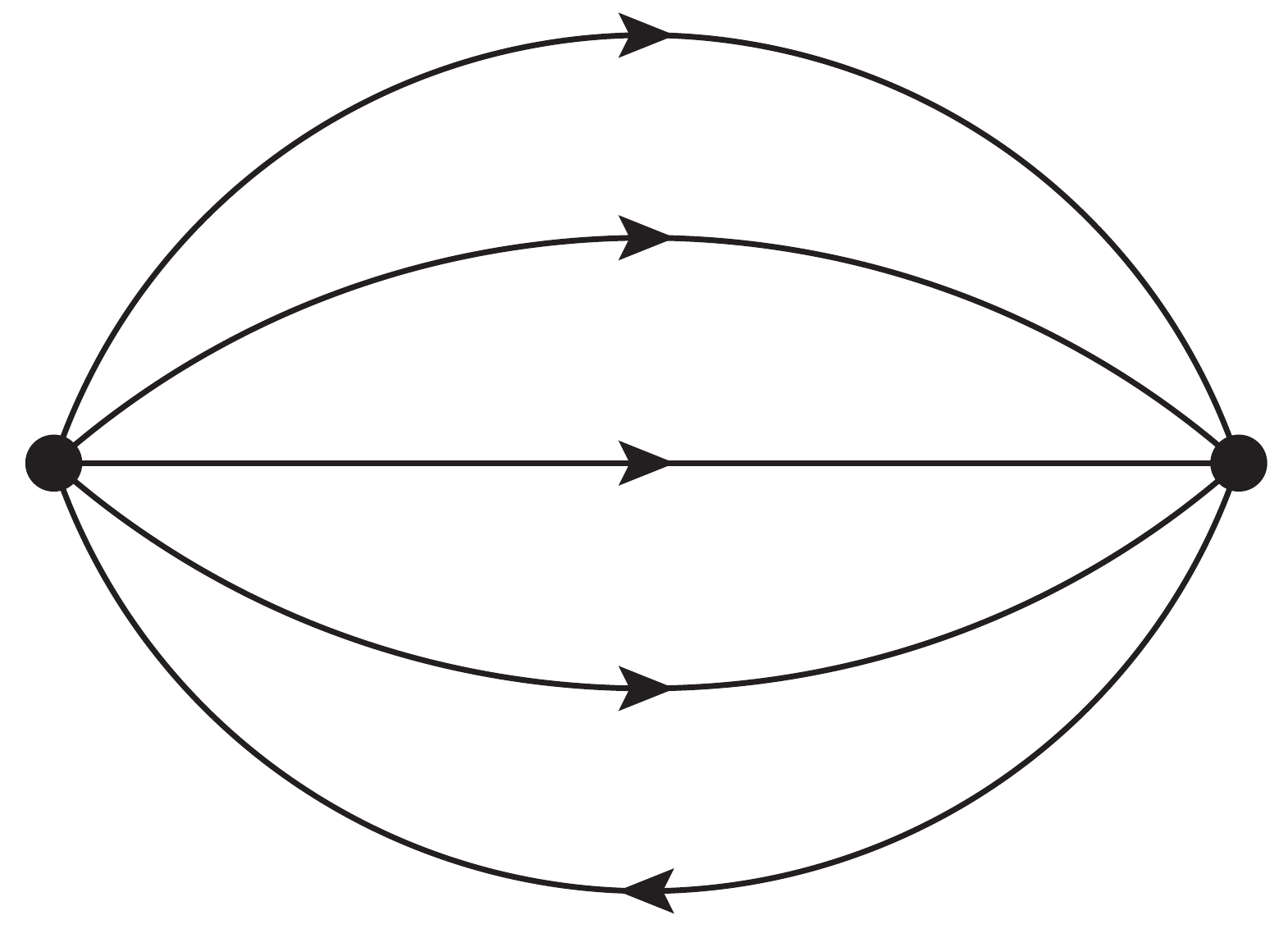}\!\!\!\!a~~~
	\includegraphics[height=0.1\textwidth]{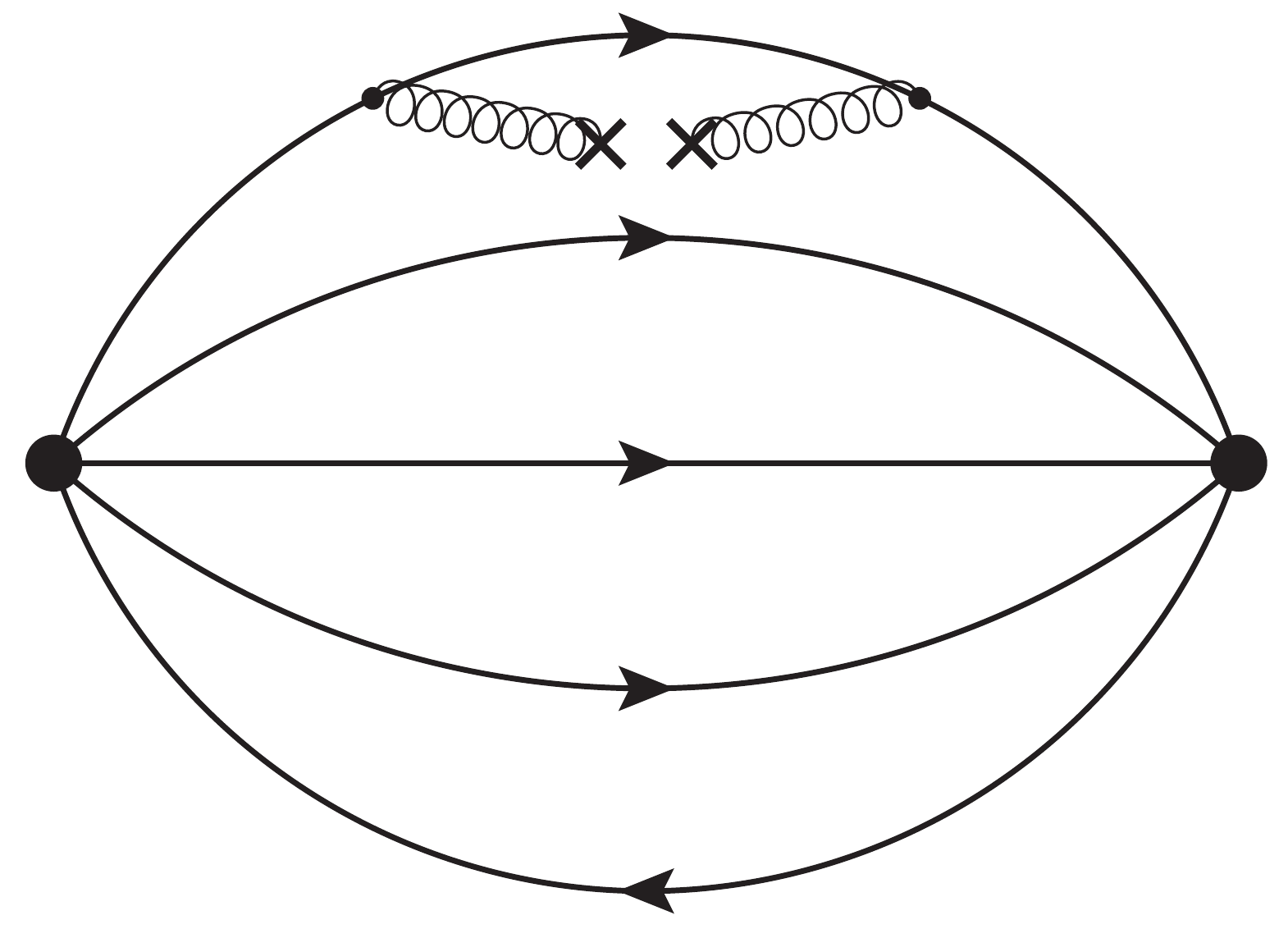}\!\!\!\!b~~~
	\includegraphics[height=0.1\textwidth]{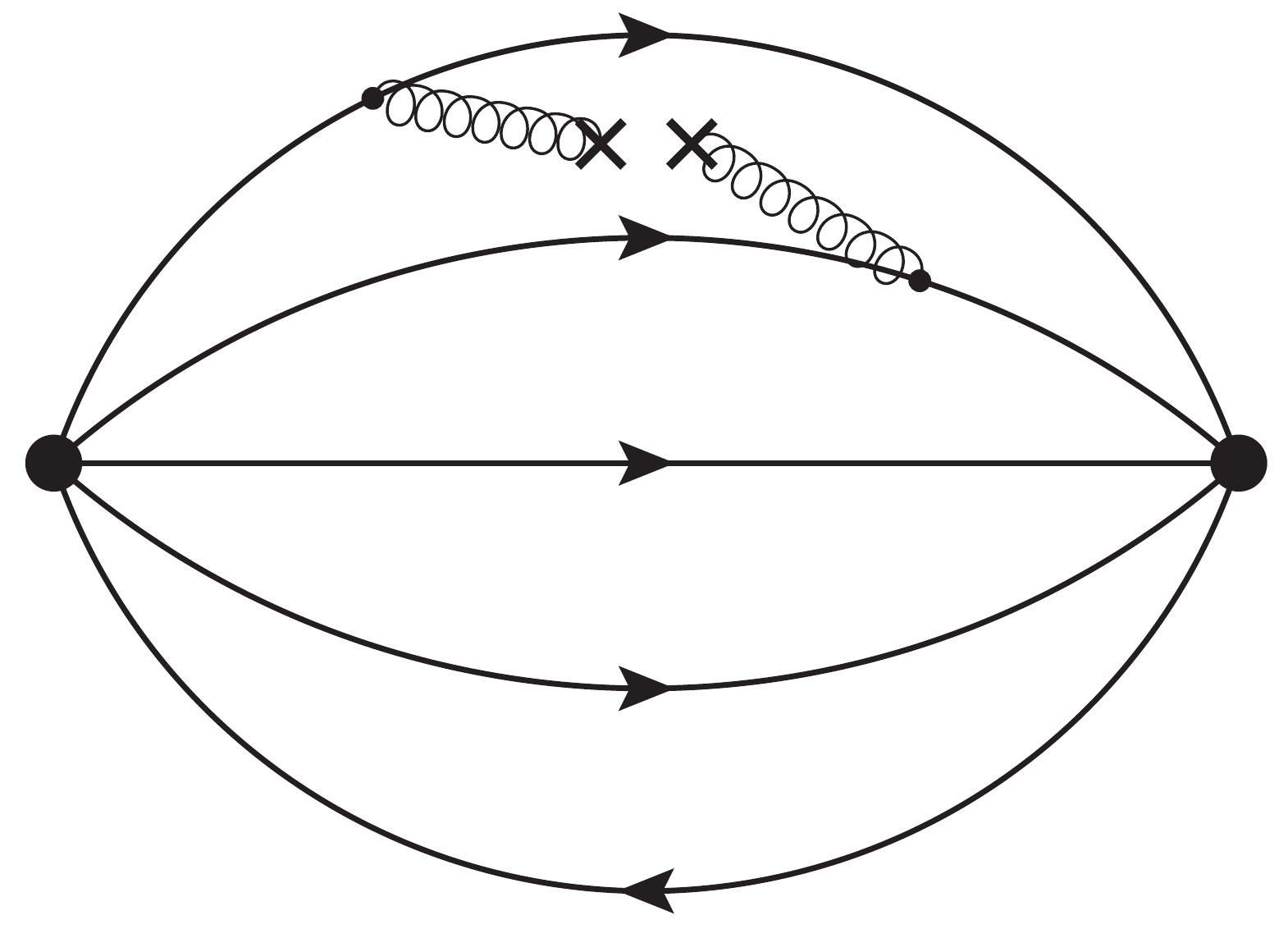}\!\!\!\!c
	\vskip 10pt
	\includegraphics[height=0.1\textwidth]{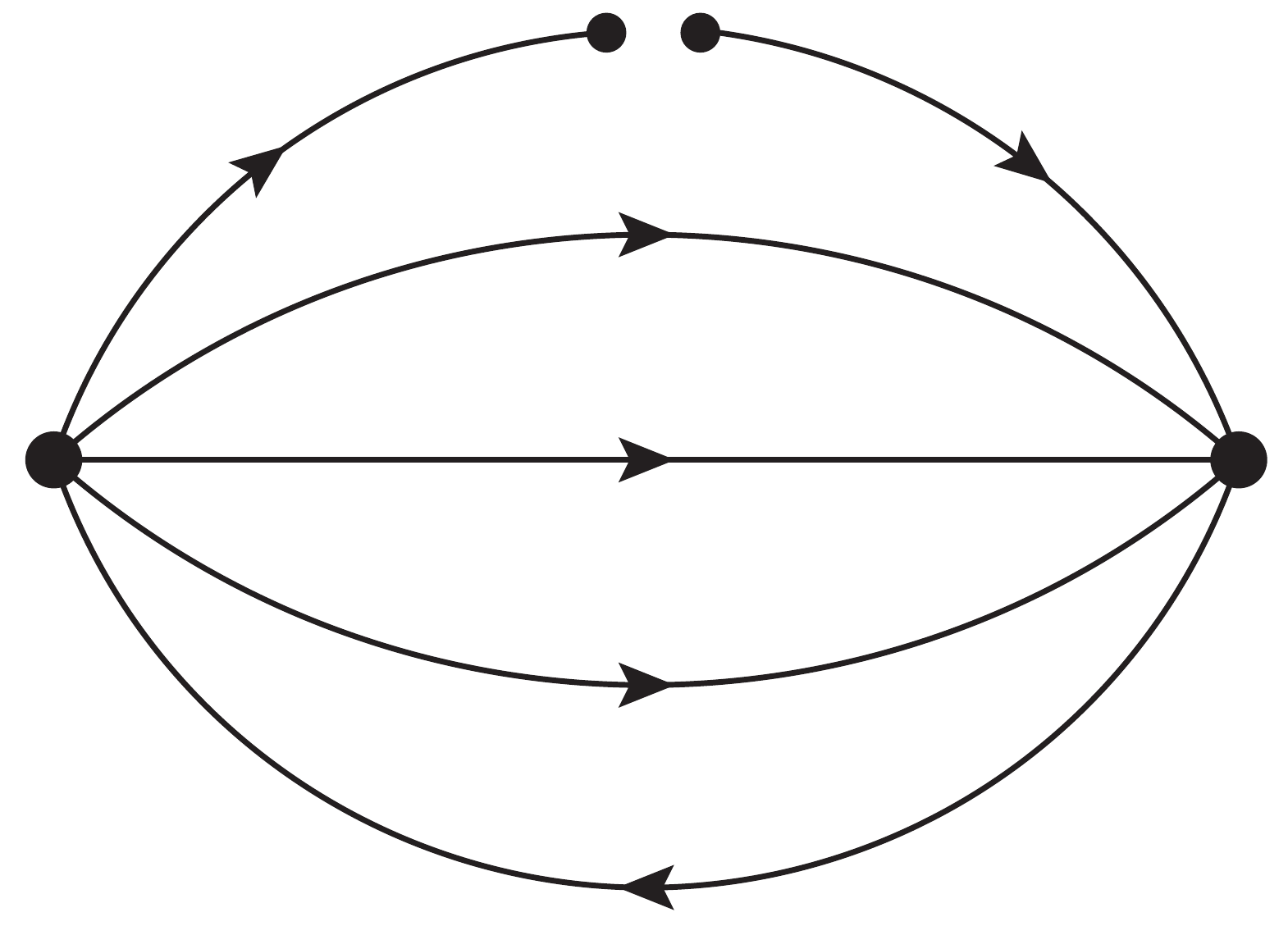}\!\!\!\!d~~~
	\includegraphics[height=0.1\textwidth]{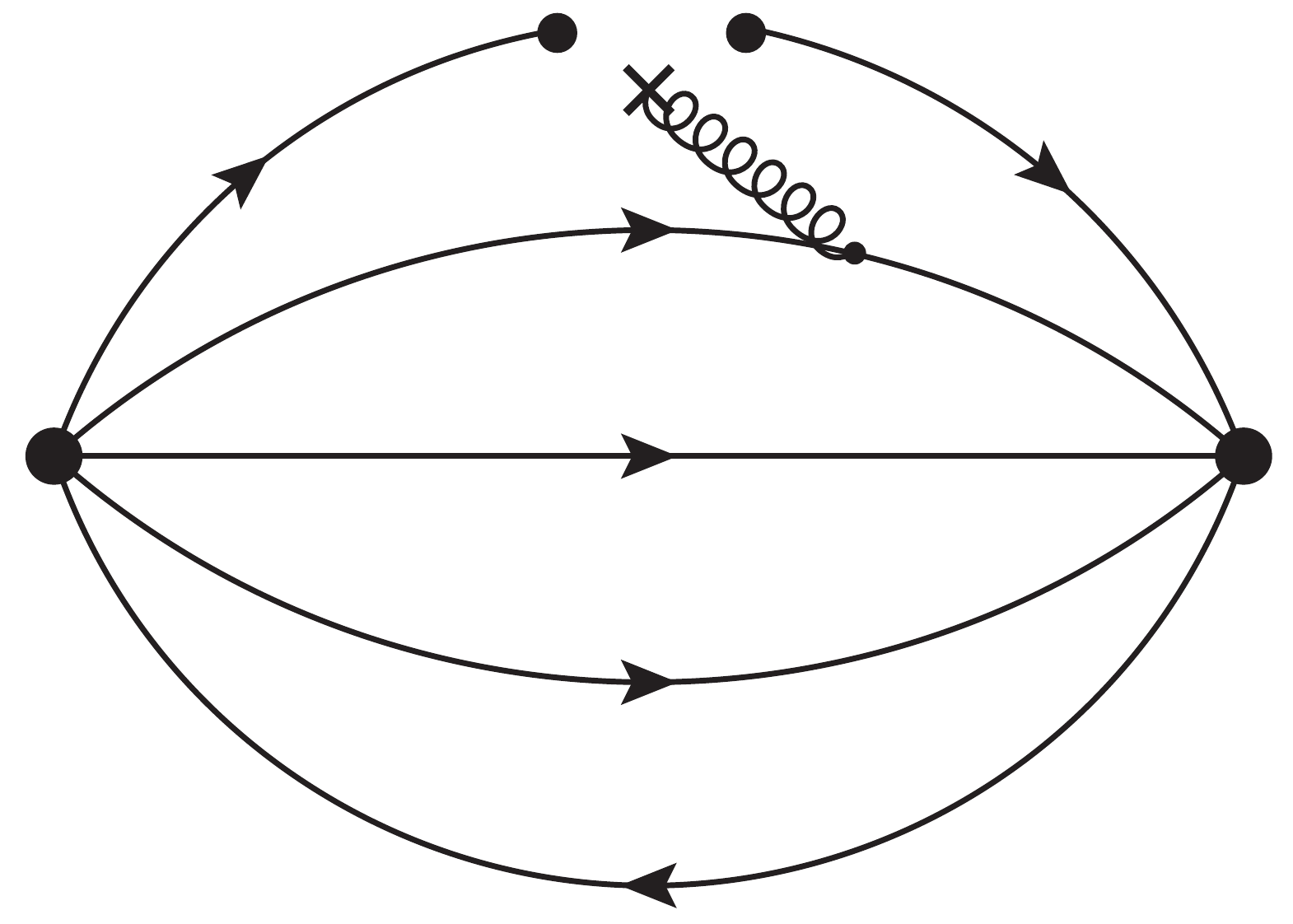}\!\!\!\!e~~~
	\includegraphics[height=0.1\textwidth]{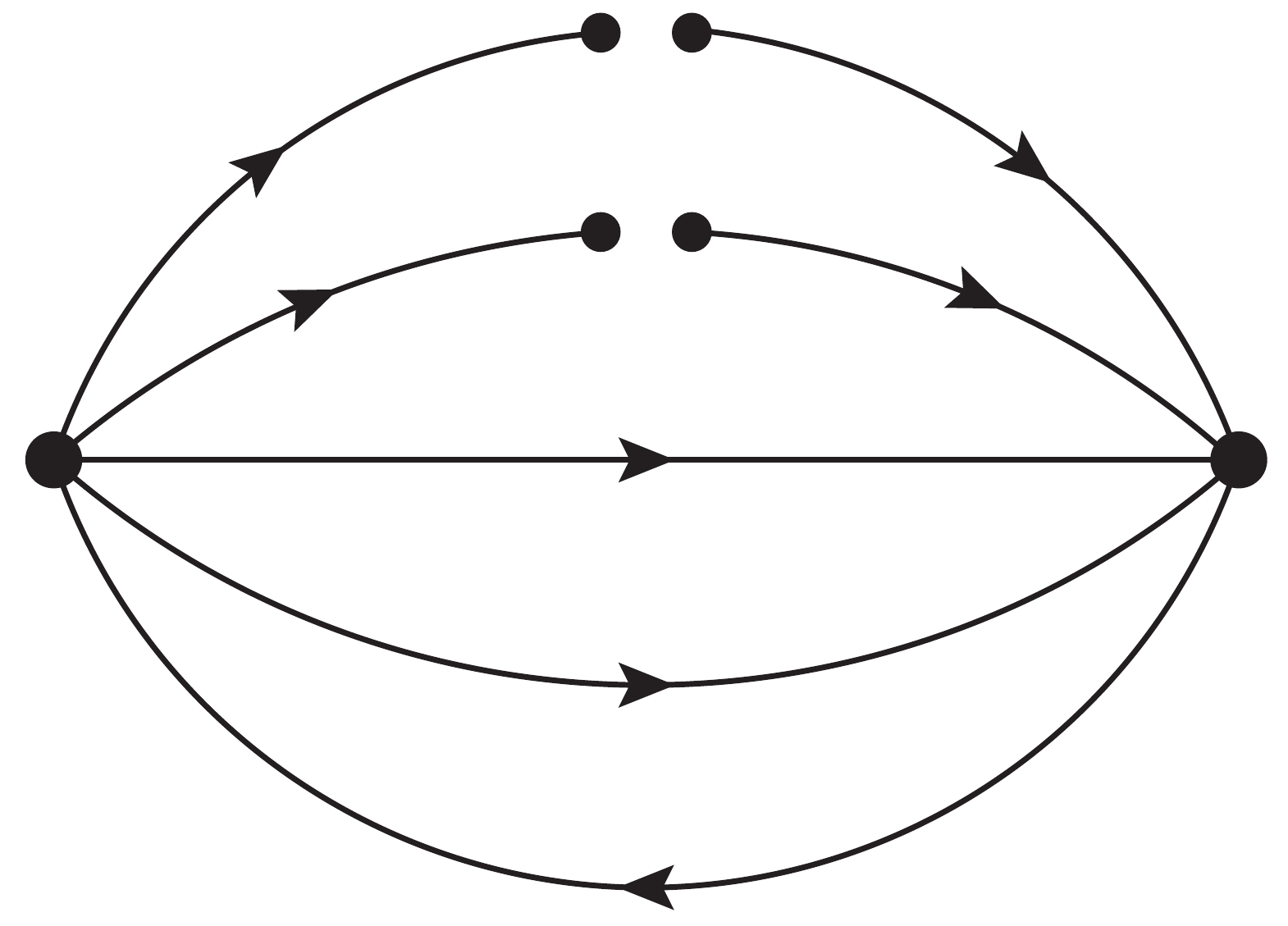}\!\!\!\!f
	\vskip 10pt
	\includegraphics[height=0.1\textwidth]{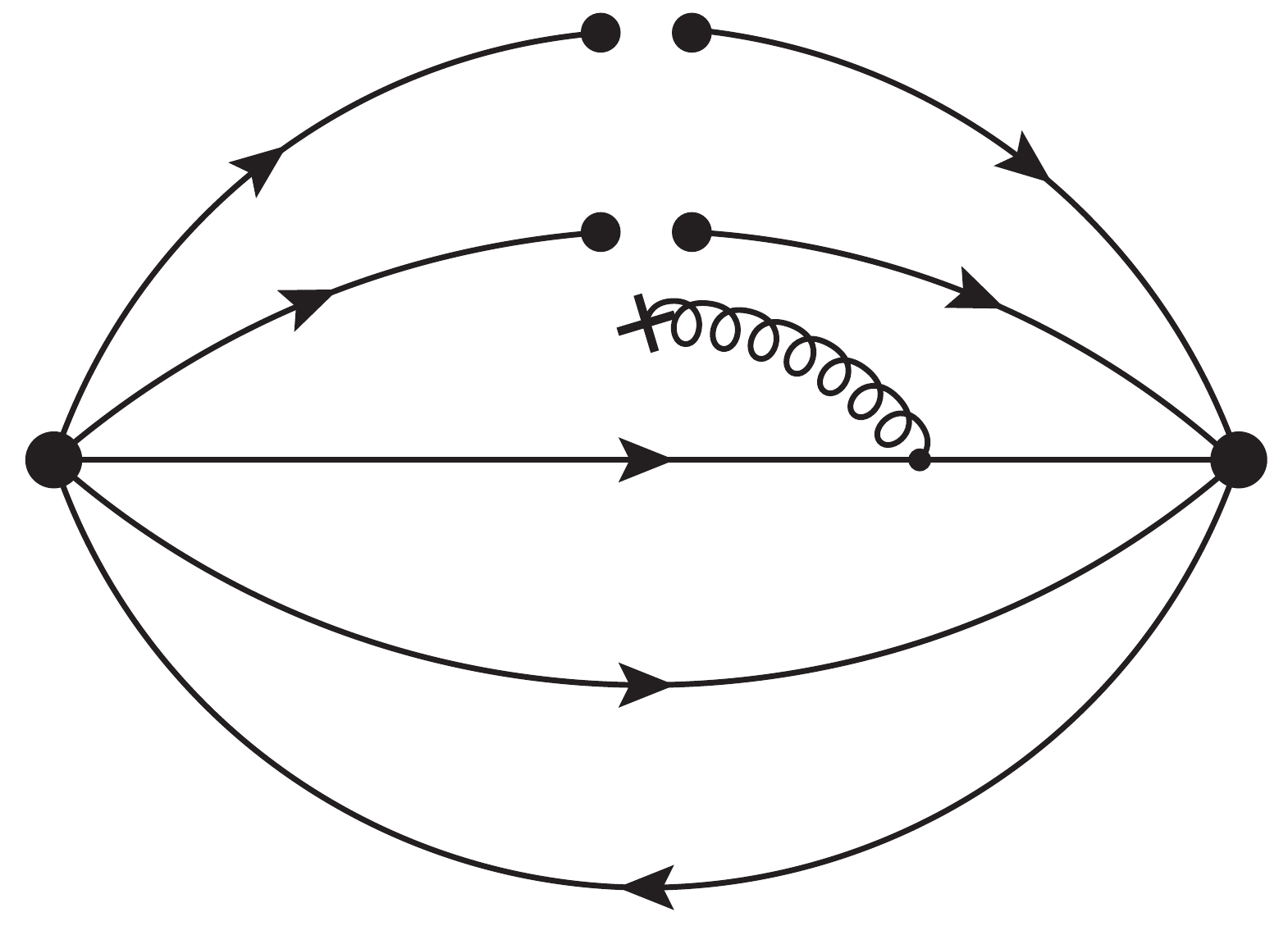}\!\!\!\!g~~~
	\includegraphics[height=0.1\textwidth]{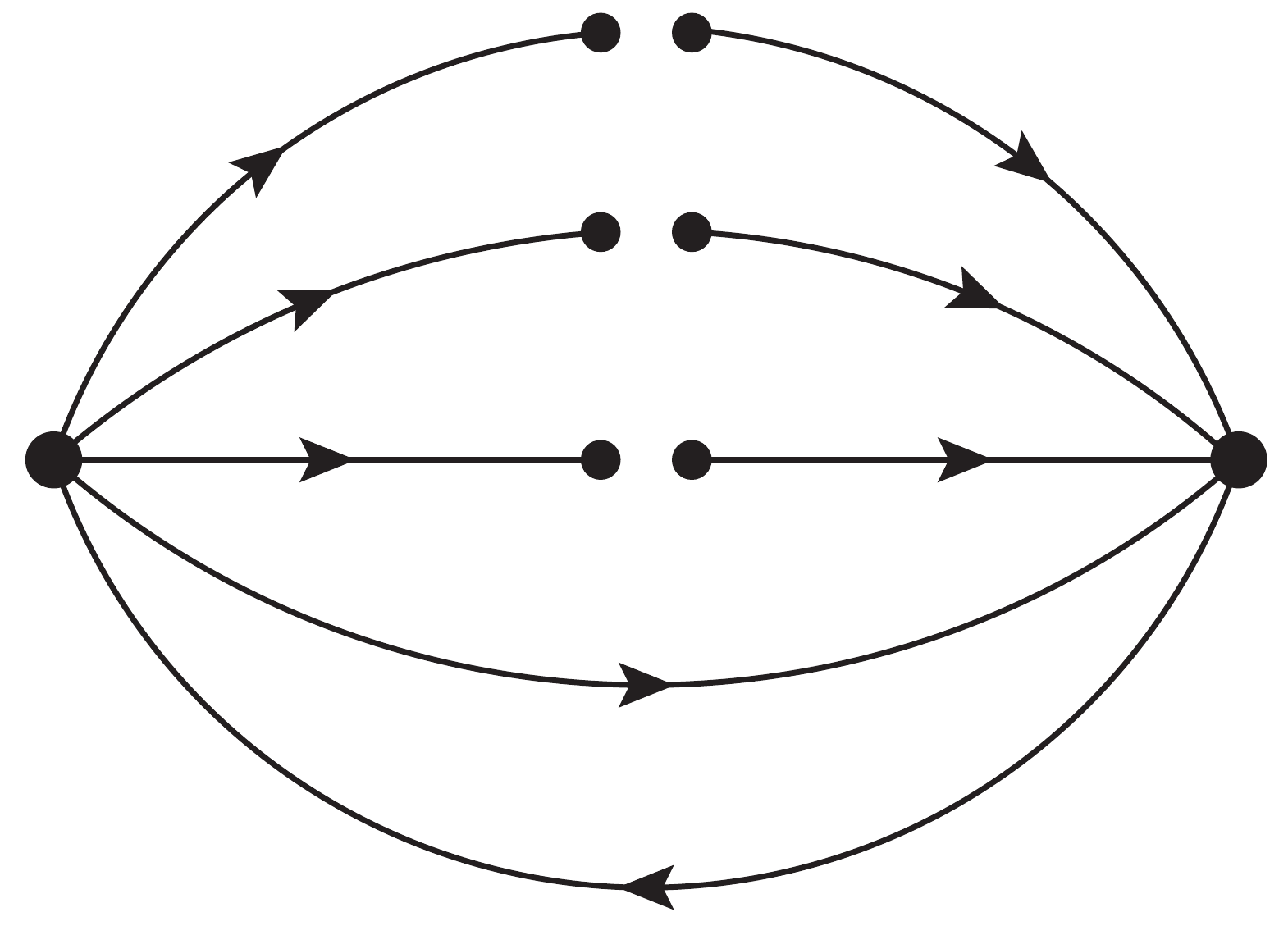}\!\!\!\!h
	\caption{
		\label{fig:ope-5quarks}
		Generic diagrams of the OPE terms for the correlators with the currents of the pentaquark states.
		Diagram (a) is the perturbative contribution at the leading order (LO). The figures (b)-(h) are
		diagrams for the nonperturbative contributions.
		We use here nonlocal condensate notation~\cite{Mikhailov:1986be,Mikhailov:1991pt,Grozin:1985wj,Grozin:1994hd,Bakulev:2006wz,Bakulev:2009ib} for
		the graphical representation of the various contributions originating
		from the standard (local) condensates.
		Some of the nonperturbative diagrams contribute to few terms of the operator OPE,
		as it is specified in Tab.~\ref{tab:condensates}.
	}
\end{figure}

In order to calculate the two functions $\Pi^s_1$ and $\Pi^s_2$
in Eq.~(\ref{eq:Pi-Sterm}) within the OPE for each current,
we use the quark propagators for both the light quarks ($u,d,s$ quarks)
and the heavy quark ($c$ quark) in the configuration space with dimension $d=4-2\epsilon$
to control ultraviolet divergences.
The heavy quark propagator in the configuration space is given by the $\al$-representation.
Our technique for the OPE calculation is similar in some aspects to that discussed in~\cite{Albuquerque:2013ija}.
We treat $u, d$ quarks as massless quarks and include the linear effect of the
strange quark mass $m_s$ in the OPE.
With the hypothesis of the vacuum dominance (HVD) factorization,
we perform the OPE up to the dimension-10 vacuum condensates so that
\begin{eqnarray}\label{eq:spectrOPE}
\rho^s_i(t)=\frac{1}{\pi}\textbf{Im}\Pi^s_i(t)=\sum\limits_{D=0}^{10}\rho^s_{i,D}(t)\,,
\end{eqnarray}
where $\rho^s_{i,D}$ is the contribution to the OPE from the dimension-$D$ condensate for each case.
The various vacuum condensates included in the OPE are listed in Tab.~\ref{tab:condensates}
with
reference to the corresponding diagrams shown in Fig.\ref{fig:ope-5quarks}.
Note, that the condensates in Tab. I are related to $u$, $d$, and $s$ quarks.
It is found that the gluon-condensate contribution is
tiny in comparison with the quark-condensate contribution.
Therefore, we don't include the contributions from the three-gluon condensate and
the dimension-7 condensate $\va{GG}\va{\bar qGq}$ to the OPE.
For the same reason, other contributions from the condensates to the OPE, which are given by the product
of the gluon condensate and the quark condensate after the HVD factorization, are also not included.
The calculated OPE contributions to the spectral density, Eq.~(\ref{eq:spectrOPE}), are given in the form of an integral
with the integrand $\rho^s_{iD}(t,\al,\be)$
\begin{eqnarray} \label{eq:rho2Dintegral}
\rho^s_{iD}(t)=
\int_{\al_-}^{\al_+}\!\!\!\!d\al\int_{\be_-}^{\be_+}\!\!\!\!d\be
\,\rho^s_{iD}(t,\al,\be)\,,
\end{eqnarray}
where the integration boundaries are
$\al_\pm=(1\pm\sqrt{1-4m_c^2/t})/2$,
$\be_+=1-\al$
and $\be_-=m_c^2\al/(t\al-m_c^2)$.

A two-dimensional integration corresponds to a two heavy-quark propagator given in the form of the $\al$-representation.
Although we consider three cases of flavor configurations
($uds$-$\bar c c$, $udc$-$\bar c s$, $usc$-${\bar c}d$),
the integrands $\rho^s_{i,k}(s,\al,\be)$ are given in Appendix D,
Eqs. (\ref{eq:rho-result-12}), (\ref{eq:rho-result-32}), (\ref{eq:rho-result-52})
only for the $uds$-$c\bar c$ configuration.

\begin{table}[h]
	\centering
	\begin{tabular}{|c|c|c|c|c|c|c|c|c|} \hline
		Term & LO & $\va{\bar qq}$ & $\va{GG}$ & $\va{\bar qGq}$ & $\va{\bar qq}^2$ &
		$\va{\bar qq}\va{\bar qGq}$  & $\va{\bar qq}^3$ & $\va{\bar qGq}^2$
		\\\hline        $D$   & 0 & 3 & 4  & 5  & 6 &  8  & 9 & 10
		\\\hline        Diag. & a & d & b, c & d, e & f &  f, g & h & f
		\\\hline
	\end{tabular}
	\caption{\label{tab:condensates}
		In the first row of the table, we list the vacuum condensates of the various operators that
		give a contribution to the OPE for the studied correlators.
		The second row provides the dimension of the operators.
		The dimension-7 condensate $\va{GG}\va{\bar qGq}$ isn't included
		in our study due to the smallness of the gluon-condensate terms.
		The third row denotes the correspondence of the operators to the diagrammatic representations
		in Fig.~\ref{fig:ope-5quarks}.
		Note that here we denote contributions from both light and s quarks
		condensates by $\va{\bar qq}^n$\,.
	}
\end{table}

\section{System of QCD SRs and Numerical Analysis}
We construct the QCD SRs for the state with spin $s$ using the
scalar functions $\Pi^s_1$ and $\Pi^s_2$ in
the correlators Eq.~(\ref{eq:Pi-Sterm}).
As discussed in the previous section, since the relativistic interpolating
current can couple to the two states with opposite parities, the physical parameters,
masses and the decay constants for the two states are coupled together in the QCD SRs.
First, we present the system of the QCD SRs in the coupled forms and discuss how to decouple the system of the QCD SRs
for each state of definite parity by using a proper combination of $\Pi^s_1$ and $\Pi^s_2$.
In this section, we omit for simplicity the index $s$ in all formulas
as far as the involved expressions are valid for any considered spin $s$.

In the framework of QCD SR~\cite{Shifman:1978bx},
the Borel transformation~$\hat{B}$
\begin{eqnarray}\nn
\hat{B}_{Q^2\to M^2}\!\left[\Pi(Q^2)\right]
= \mathop{\text{lim}}\limits_{n\to\infty}\!
\frac{(-Q^2)^n}{\Gamma(n)}\!
\left[\frac{d^n}{dQ^{2n}}\Pi(Q^2)\right]_{Q^2=n M^2}\,,
\end{eqnarray}
is applied to both sides of Eq.~(\ref{eq:SR-disp}).
This transformation helps to reduce the SR uncertainties by suppressing the contributions
from the excited resonances in the continuum and also higher-order OPE terms.

For the phenomenological part of the SR, we apply the phenomenological spectral densities, which are called by $\rho_i^\text{ph}(t)$ and appear on the right-hand side in Eq.~(\ref{eq:SR-disp}).
For all considered states, we assume that these spectral densities
can be decomposed into contributions from the resonances of the considered states
and the contribution from the continuum starting from the threshold $s_0$ appealing to the quark-hadron duality hypothesis
\begin{eqnarray}\nn
\rho_1^\text{ph}(t)&=&
f^2_+\delta(t-m^2_+)+
f^2_-\delta(t-m^2_-)\\\nn
&&+\Theta(t-s_0)\rho^\text{OPE}_1(t)\,,\\\nn
\rho_2^\text{ph}(t)&=&
f^2_+m_+\delta(t-m^2_+)-
f^2_-m_-\delta(t-m^2_-)\\\nn
&&+\Theta(t-s_0)\rho^\text{OPE}_2(t)\, ,
\end{eqnarray}
where the threshold $s_0$ is chosen to be the same for both
parities and for both densities ($\rho_1^\text{ph}$ and $\rho_2^\text{ph}$).
The OPE spectral densities $\rho^\text{OPE}_i(t)=\rho^s_i(t)$
are defined by Eq.~(\ref{eq:spectrOPE}).
The decay constants $f_\pm$ and masses $m_\pm$ are given in
Eqs.~(\ref{eq:spinSum12}), (\ref{eq:spinSum32}), (\ref{eq:spinSum52}).
Then, the resonance contributions to the phenomenological part of the SR are defined as follows
\begin{eqnarray}\nn
{\cal R}^\text{(res)}_{1,k}(M^2)&=&f_{+}^2 m_{+}^{2k}e^{-m_{+}^2/M^2}
+ f_{-}^2 m_{-}^{2k}e^{-m_{-}^2/M^2}\,,\\\nn
{\cal R}^\text{(res)}_{2,k}(M^2)&=&f_{+}^2 m_{+}^{2k+1}e^{-m_{+}^2/M^2}
- f_{-}^2 m_{-}^{2k+1}e^{-m_{-}^2/M^2}\,,
\end{eqnarray}
where we apply the Borel transformation to Eq.~(\ref{eq:SR-disp}), as already discussed.
Combining the full OPE results with the contribution from the continuum,
we evaluate the theoretical part of the QCD SRs
\begin{eqnarray}\nn
{\cal R}^\text{(SR)}_{i,k}(M^2,s_0)=
\int_{s_\text{th}}^{s_0}\!dt\,\rho_i(t)\,t^ke^{-t/M^2}\,,
\end{eqnarray}
where the $k$-times derivatives with respect to $-1/M^2$ are taken after the Borel transformation.
Finally, for each state of spin $s$=1/2, 3/2, 5/2, we obtain the
following system of QCD SRs
in the coupled form:
\begin{eqnarray}\label{eq:SR12}
{\cal R}^\text{(res)}_{1,k}(M^2)={\cal R}^\text{(SR)}_{1,k}(M^2,s_0)\,,\\\nn
{\cal R}^\text{(res)}_{2,k}(M^2)={\cal R}^\text{(SR)}_{2,k}(M^2,s_0)\,.
\end{eqnarray}
where $k\in Z_+\bigcup\{0\}$.

\begin{figure*}[t]
    \includegraphics[height=0.2\textwidth]{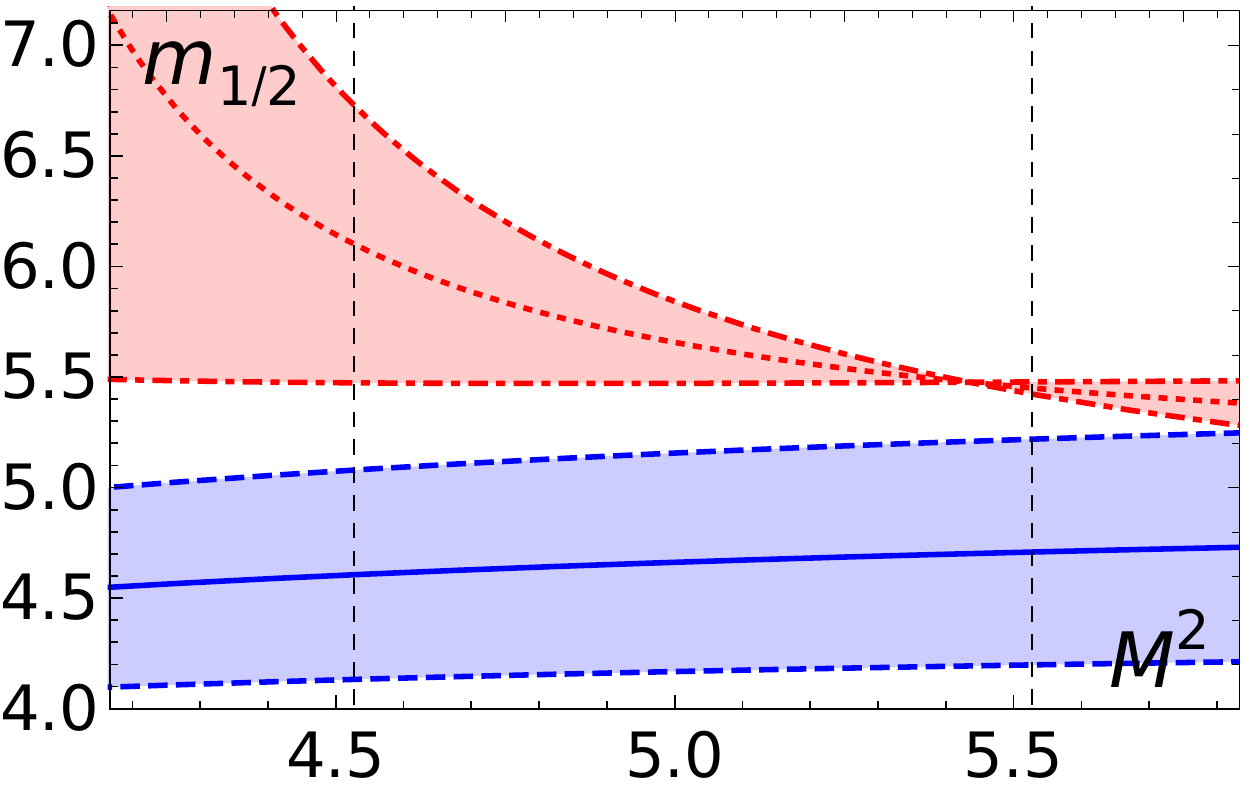}~~~
    \includegraphics[height=0.2\textwidth]{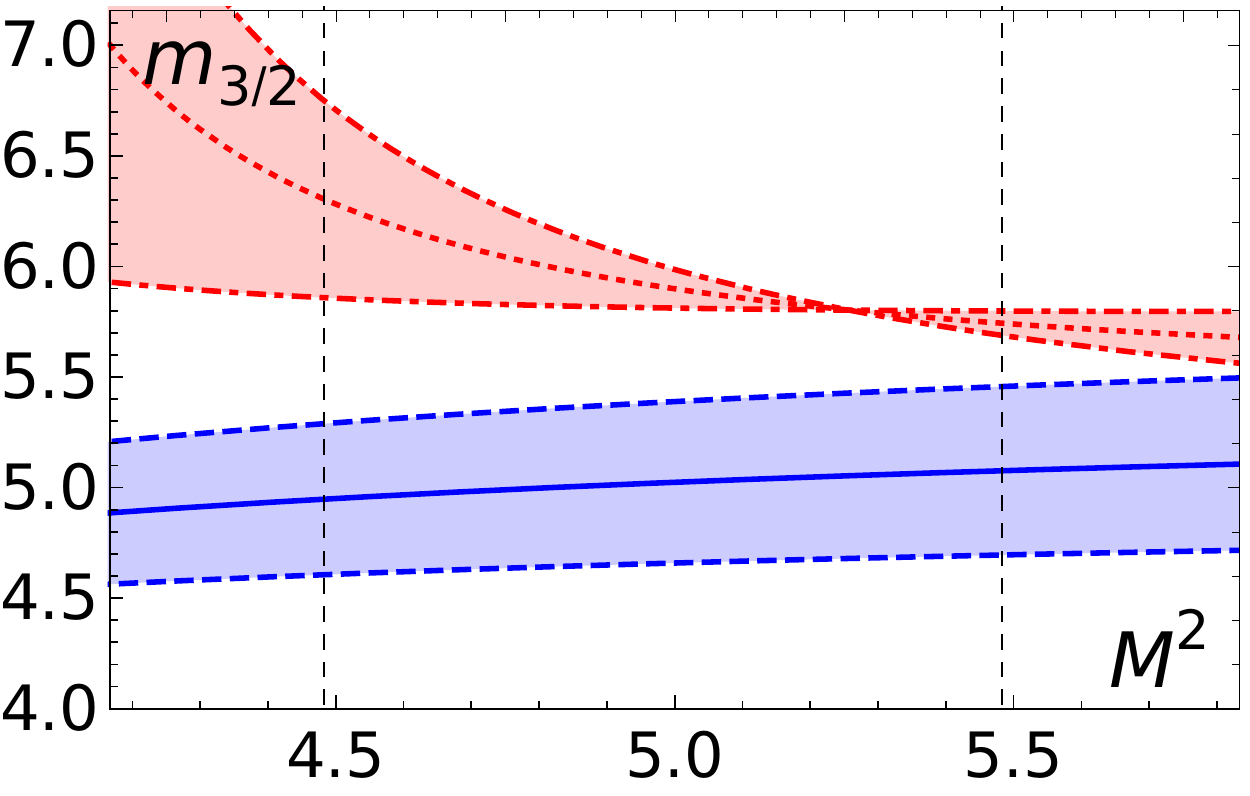}~~~
    \includegraphics[height=0.2\textwidth]{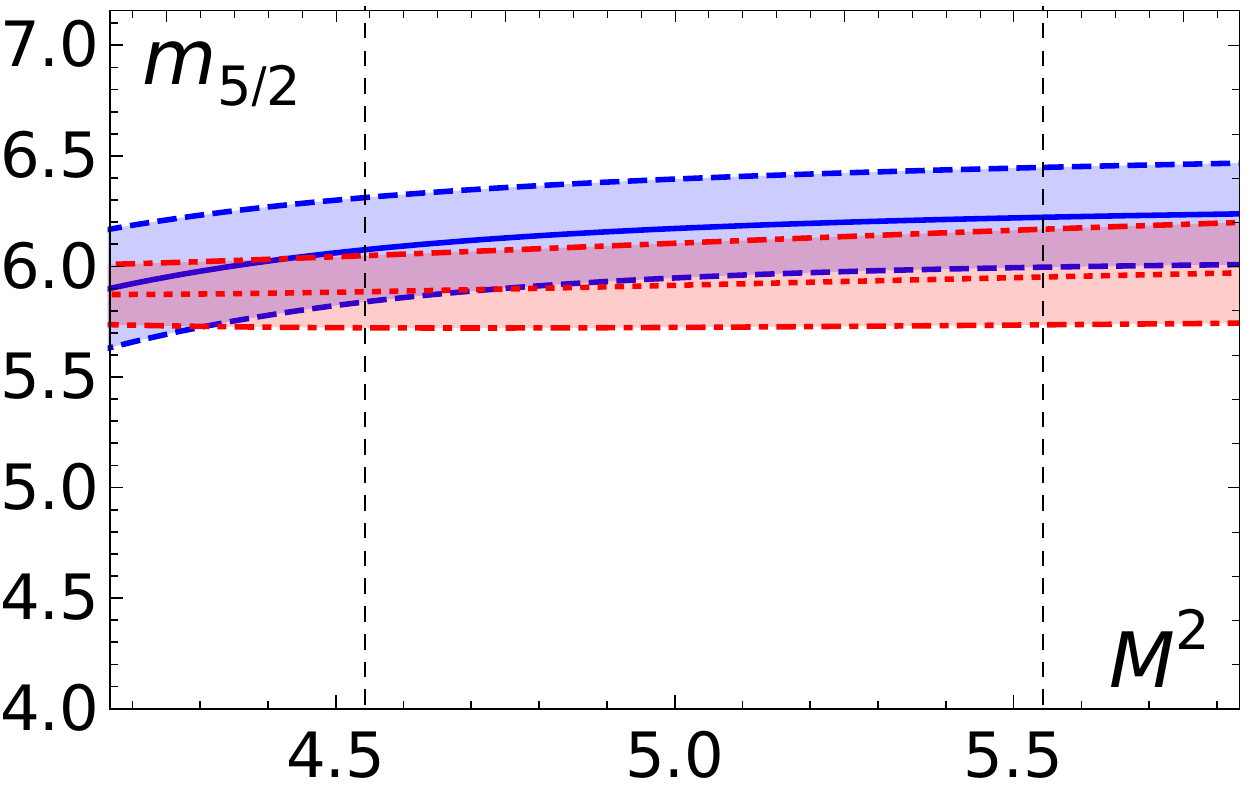}
    \caption{
        \label{fig:fig-SR}
Borel parameter dependence of the mass $m_s(s_0,M^2)$
for a $uds$-$\bar cc$ flavor clustering, given by Eq.~(\ref{eq:massSRborel}),
referring to
 spin $s=1/2$ (left panel),
 spin $s=3/2$ (central panel),
 spin $s=5/2$ (right panel) for
 negative parity (blue solid line and blue band limited by dashed blue lines) and
 positive parity (red dotted line and red band limited by dot-dashed red lines).
The central lines of the bands denote the dependence
for the best fit threshold $s_0=\tilde s_0$.
The bands show the dependence of the masses on the threshold $s_0$
varied in the threshold interval $s_0\in[s_0^\text{min},s_0^\text{max}]$.
Vertical dotted black lines present the Borel windows $(M^2_-,M^2_+)$.
    }
\end{figure*}

\subsection{Decoupled QCD SRs}

This subsection is devoted to decoupling the SRs in Eqs.~(\ref{eq:SR12}) 
into two QCD SR equations for each state of definite parity.
It seems that there are four different ways to deal with this kind of coupled QCD SRs systems
used in the pentaquark QCD SR studies.
First, assuming that most of the contributions come from the
lowest lying resonance of the considered parity, the contributions from the resonance
of the opposite parity can be ignored and
only the second equation in Eq.~(\ref{eq:SR12}) has been considered.
This approach has been applied to many studies on the states of $S=1/2$
and to pentaquark states~\cite{Lee:2005ny,Xiang:2017byz,Chen:2016qju}.
In a second way, used in~\cite{Azizi:2016dhy}, one resolves the systems (\ref{eq:SR12})
by taking into account the states of both parities
without decoupling the system.
The third way is to get the decoupled QCD SRs by using the old-fashioned correlator~\cite{Jido:1996ia,Ohtani:2012ps}.
Here, we use a method that is similar to the fourth way~\cite{Wang:2015epa}, in which
the system of SRs, see, Eq.~(\ref{eq:SR12}), is decoupled into two QCD SRs for each state
of definite parity.

To decouple the SRs given by Eqs.~(\ref{eq:SR12}), we
expand the region of validity for k to ${k\in\{n/2|n\in Z\}}$.
This analytical continuation allows us to consider
the following linear combination of Eqs.~(\ref{eq:SR12})
\begin{eqnarray}\nn
{\cal R}^\text{(SR)}_{\pm,k}=\frac 12 (
{\cal R}^\text{(SR)}_{1,k} \pm
{\cal R}^\text{(SR)}_{2,k-1/2})\,,\\\nn
{\cal R}^\text{(res)}_{\pm,k}=\frac 12 (
{\cal R}^\text{(res)}_{1,k} \pm
{\cal R}^\text{(res)}_{2,k-1/2})\,.
\end{eqnarray}
As a result we can rewrite the SRs, given by Eq.~(\ref{eq:SR12}), in decoupled form to read
\begin{eqnarray}\label{eq:SRparity}
{\cal R}^\text{(res)}_{\pm,k}(M^2)&=&
{\cal R}^\text{(SR)}_{\pm,k}(M^2,s_0)\,,
\end{eqnarray}
with
\begin{eqnarray}\nn
{\cal R}^\text{(res)}_{\pm,k}(M^2)&=&f^2_\pm e^{-m^2_\pm/M^2}m^{2k}_\pm\,,\\\nn
{\cal R}^\text{(SR)}_{\pm,k}(M^2,s_0)&=&\int_{s_\text{th}}^{s_\pm}dt
\rho_\pm^\text{OPE}(t) t^k e^{-t/M^2}\,.
\end{eqnarray}
where the reparameterized spectral densities $\rho_{\pm}^\text{OPE}$ are related to $\rho_{1,2}^\text{OPE}$
(calculated by the OPE in Eq.~(\ref{eq:density})) as
\begin{equation}\nn
\rho^\text{OPE}_{\pm}(t)=\frac{1}{2}\bigg(\rho^\text{OPE}_1(t)
\pm\frac{\rho^\text{OPE}_2(t)}{\sqrt{t}}\bigg)\ .
\end{equation}
The decoupled QCD SRs, Eq.~(\ref{eq:SRparity}), can be written in explicit form
\begin{eqnarray}\label{eq:SRexplicit}
f^2_\pm e^{-m^2_\pm/M^2}m^{2k}_\pm = \int_{s_\text{th}}^{s_\pm}dt
\rho_\pm(t) t^k
e^{-t/M^2}\,.
\end{eqnarray}

\begin{table*}[t]
    \centering
    $$
    \begin{array}{|c|c||c|c|c|c|c|c|c|c|c|} \hline
    \text{spin} &  \text{flavor}
    & m_-(\tilde s_0),\text{GeV}
    & m_+(\tilde s_0),\text{GeV}
    & 10^3f_-(\tilde s_0),\text{GeV}^6
    & 10^3f_+(\tilde s_0),\text{GeV}^6
    & (s_0^\text{min},s_0^\text{max})
    & \tilde s_0
    & (M^2_-,M^2_+)
    & \delta(\tilde s_0)\,,\%
    & r(\tilde s_0) \\ \hline\hline
    1/2 & uds\text{-}\bar cc & 4.4_{-0.3}^{+0.7} \pm 0.0 & 5.1_{-0.0}^{+1.0} \pm 0.1 & 1.8_{-0.7}^{+4.0} \pm 0.0 & 1.3_{-0.2}^{+3.9} \pm 0.0 & (22.0,34.7)  &  25.1 & (4.5 , 5.5)  &  3.8 & 0.05 \\ \hline
    1/2 & udc\text{-}\bar cs & 4.5_{-0.2}^{+0.3} \pm 0.1 & 5.3_{-0.0}^{+0.3} \pm 0.2 & 2.0_{-0.5}^{+1.4} \pm 0.1 & 2.0_{-0.1}^{+1.4} \pm 0.0 & (24.0,30.6)  &  26.1 & (3.7 , 4.7)  &  5.0 & 0.13 \\ \hline
    1/2 & usc\text{-}\bar cd & 4.6_{-0.2}^{+0.3} \pm 0.1 & 5.3_{-0.0}^{+0.3} \pm 0.2 & 2.1_{-0.6}^{+1.6} \pm 0.1 & 2.1_{-0.2}^{+1.6} \pm 0.1 & (24.3,31.2)  &  26.5 & (3.7 , 4.7)  &  5.1 & 0.14 \\ \hline
    3/2 & uds\text{-}\bar cc & 4.9_{-0.2}^{+0.5} \pm 0.1 & 5.7_{-0.0}^{+0.6} \pm 0.2 & 1.9_{-0.6}^{+2.5} \pm 0.0 & 1.9_{-0.2}^{+2.6} \pm 0.0 & (27.3,38.6)  &  30.2 & (4.5 , 5.5)  &  4.7 & 0.13 \\ \hline
    3/2 & udc\text{-}\bar cs & 4.8_{-0.1}^{+0.2} \pm 0.1 & 5.8_{-0.0}^{+0.2} \pm 0.3 & 1.9_{-0.4}^{+0.8} \pm 0.1 & 2.6_{-0.0}^{+0.8} \pm 0.1 & (28.7,34.9)  &  30.8 & (3.4 , 4.4)  &  7.4 & 0.32 \\ \hline
    3/2 & usc\text{-}\bar cd & 4.9_{-0.1}^{+0.2} \pm 0.1 & 5.8_{-0.0}^{+0.2} \pm 0.3 & 2.1_{-0.4}^{+0.8} \pm 0.1 & 3.0_{-0.0}^{+0.9} \pm 0.1 & (29.5,36.1)  &  31.8 & (3.4 , 4.4)  &  8.2 & 0.36 \\ \hline
    5/2 & uds\text{-}\bar cc & 6.2_{-0.3}^{+0.1} \pm 0.1 & 6.0_{-0.3}^{+0.1} \pm 0.0 & 12.1_{-6.4}^{+3.8} \pm 0.1 & 15.6_{-6.1}^{+3.3} \pm 0.1 & (39.2,50.0)  &  46.3 & (4.5 , 5.5)  &  1.2 & 0.51 \\ \hline
    5/2 & udc\text{-}\bar cs & 6.0_{-0.4}^{+0.1} \pm 0.1 & 5.9_{-0.3}^{+0.1} \pm 0.0 & 7.2_{-4.1}^{+2.5} \pm 0.1 & 12.2_{-4.9}^{+2.6} \pm 0.1 & (37.6,50.0)  &  45.5 & (3.7 , 4.7)  &  1.7 & 0.64 \\ \hline
    5/2 & usc\text{-}\bar cd & 6.3_{-0.3}^{+0.0} \pm 0.1 & 6.0_{-0.3}^{+0.0} \pm 0.1 & 9.9_{-5.3}^{+0.0} \pm 0.1 & 15.1_{-5.9}^{+0.0} \pm 0.2 & (40.6,50.0)  &  50.0 & (3.8 , 4.8)  &  2.1 & 0.75
    \\ \hline
    \end{array}
    $$
    \caption{\label{tab:SR-results}
        QCD SR results for masses $m_\pm$ and decay constants $f_\pm$ given for a pentaquark with both parities
        with spin 1/2, 3/2, 5/2 (first column) for three cases of flavor-clustering (second column).
        For all cases considered in this table, we apply the type-1 currents defined in Eqs.
        (\ref{eq:current12types}),
        (\ref{eq:current32types}),
        (\ref{eq:current52types}).
        Central values of masses (2nd and 3rd columns)
        and decay constants (4th and 5th columns) given at
        the best-fit threshold $(\tilde s_0)$ (see column 8th).
        The first error bars from the third to the sixth column represent
        the variation with respect to the threshold value in the interval
        $(s_0^\text{min},s_0^\text{max})$ given in the 7th column.
        The second error bars in the columns from the third to the sixth represent
        the variation in the Borel window $(M^2_-,M^2_+)$ (see 9th column).
        The criteria values $\delta(\tilde s_0)$ are given for each state in percentages in the 10th column.
        Additionally, the last column represents the criteria of the resonance contribution $r_1(\tilde s_0)$.
    }
\end{table*}

\subsection{Numerical Analysis}
In this subsection, we extract the masses and the decay constants from the constructed QCD SRs. The first step is to define the Borel window $M^2\in[M^2_-,M^2_+]$
by the conditions
\begin{eqnarray}\nn
\frac{{\cal R}^\text{(SR)}_{2,9,0}(M^2_-,\infty)}
{{\cal R}^\text{(SR)}_{2,0}(M^2_-,\infty)}<\frac{1}{10}\,,~~
M^2_+=M^2_-+\Delta M^2\,.
\end{eqnarray}
The low boundary $M^2_-$ of the Borel window insures that the dimension-9 condensate $\va{\bar qq}^3$
contributes less than 10\% to the total value of the correlator.
Here we use the following notation for the OPE contribution of dimension D
\begin{eqnarray}\nn
{\cal R}^\text{(SR)}_{i,D,k}(M^2,s_0)=
\int_{s_\text{th}}^{s_0}\!dt\,\rho_{iD}(t)\,t^ke^{-t/M^2}\,.
\end{eqnarray}
The upper boundary $M^2_+$ is determined by the above condition by setting $\Delta M^2=1$~GeV$^2$.
We don't follow the common practice to define the upper boundary $M^2_+$ by the condition
that the resonance contribution gives at least 10\%
to the total value of the correlator, $r_i(s_0)>1/10$,
for $i=1\,,2$, where
\begin{eqnarray}\nn 
r_i(s_0) =
\frac{{\cal R}^\text{(SR)}_{i,0}(M^2_+, s_0)}
{{\cal R}^\text{(SR)}_{i,0}(M^2_+,\infty)}\,.
\end{eqnarray}
The values of this ratio are given in Tables~\ref{tab:SR-results},
\ref{tab:SR-results-udcsC}, and
\ref{tab:SR-results-udcuC} for the considered SRs.
Note that most of the SRs yield values of this ratio above 1/10.
Having an equal size of the Borel window $\Delta M^2$ for all SRs allows us to compare the SR stability criteria for different SRs
without violating the condition $r_i(s_0)>1/10$.
To control this condition we introduce the collective value
\begin{eqnarray}\nn
r(s_0)=\textbf{min}(r_{1}(s_0),r_{2}(s_0))\,,
\end{eqnarray}
that can be found in the last column of Tables~\ref{tab:SR-results},
\ref{tab:SR-results-udcsC}, and
\ref{tab:SR-results-udcuC}.

\begin{table*}[t]
    \centering
    $$
    \begin{array}{|c|c||c|c|c|c|c|c|c|c|c|} \hline
    \text{spin} &  \text{type}
    & m_-(\tilde s_0),\text{GeV}
    & m_+(\tilde s_0),\text{GeV}
    & 10^3f_-(\tilde s_0),\text{GeV}^6
    & 10^3f_+(\tilde s_0),\text{GeV}^6
    & (s_0^\text{min},s_0^\text{max})
    & \tilde s_0
    & (M^2_-,M^2_+)
    & \delta(\tilde s_0)\,,\%
    & r(\tilde s_0) \\ \hline\hline
    1/2 & 1 (1) & 4.4_{-0.3}^{+0.7} \pm 0.0 & 5.1_{-0.0}^{+1.0} \pm 0.1 & 1.8_{-0.7}^{+4.0} \pm 0.0 & 1.3_{-0.2}^{+3.9} \pm 0.0 & (22.0,34.7)  &  25.1 & (4.5 , 5.5)  &  3.8 & 0.05 \\ \hline
    1/2 & 2(1) & 4.4_{-0.3}^{+0.9} \pm 0.0 & 5.1_{-0.0}^{+1.0} \pm 0.1 & 1.8_{-0.7}^{+5.6} \pm 0.0 & 1.3_{-0.2}^{+5.4} \pm 0.0 & (22.0,36.6)  &  25.1 & (4.6 , 5.6)  &  4.2 & 0.04 \\ \hline
    1/2 & 3(0) & 5.9_{-0.0}^{+0.9} \pm 0.2 & 5.1_{-0.2}^{+0.6} \pm 0.0 & 1.4_{-0.1}^{+2.8} \pm 0.0 & 1.6_{-0.5}^{+2.7} \pm 0.0 & (29.6,43.1)  &  32.7 & (4.9 , 5.9)  &  4.0 & 0.1 \\ \hline
    1/2 & 4(0) & 5.9_{-0.0}^{+0.9} \pm 0.2 & 5.1_{-0.2}^{+0.6} \pm 0.0 & 4.2_{-0.3}^{+8.3} \pm 0.1 & 4.7_{-1.5}^{+8.1} \pm 0.1 & (29.6,43.1)  &  32.7 & (4.9 , 5.9)  &  4.0 & 0.1 \\ \hline
    3/2 & 1(1) & 4.9_{-0.2}^{+0.5} \pm 0.1 & 5.7_{-0.0}^{+0.6} \pm 0.2 & 1.9_{-0.6}^{+2.5} \pm 0.0 & 1.9_{-0.2}^{+2.6} \pm 0.0 & (27.3,38.6)  &  30.2 & (4.5 , 5.5)  &  4.7 & 0.13 \\ \hline
    3/2 & 2(0) & 6.2_{-0.4}^{+0.1} \pm 0.0 & 6.6_{-0.1}^{+0.2} \pm 0.0 & 6.0_{-3.1}^{+0.9} \pm 0.0 & 4.9_{-2.7}^{+1.0} \pm 0.0 & (38.8,50.0)  &  47.8 & (4.9 , 5.9)  &  0.8 & 0.4
    \\ \hline
    \end{array}
    $$
    \caption{\label{tab:SR-results-udcsC}
        QCD SR results for the masses $m_\pm$ and the decay constants $f_\pm$ given for a pentaquark of both parities
        with spin 1/2, 3/2 (first column) and for a
        $udc$-$\bar cs$ flavor-clustering.
        The second column denotes the type of the current and the spin of the $\bar cs$-part given in the parentheses.
        The types of the currents are defined in Eqs.~(\ref{eq:current12types}) and
        (\ref{eq:current32types}).
        See the caption of Table~\ref{tab:SR-results} for more details.
    }
\end{table*}

\begin{table*}[t]
    \centering
    $$
    \begin{array}{|c|c||c|c|c|c|c|c|c|c|c|} \hline
    \text{spin} &  \text{type}
    & m_-(\tilde s_0),\text{GeV}
    & m_+(\tilde s_0),\text{GeV}
    & 10^3f_-(\tilde s_0),\text{GeV}^6
    & 10^3f_+(\tilde s_0),\text{GeV}^6
    & (s_0^\text{min},s_0^\text{max})
    & \tilde s_0
    & (M^2_-,M^2_+)
    & \delta(\tilde s_0)\,,\%
    & r(\tilde s_0) \\ \hline\hline
1/2 & 1 & 4.4_{-0.2}^{+0.4} \pm 0.0 & 5.1_{-0.0}^{+0.4} \pm 0.1 & 1.6_{-0.5}^{+1.5} \pm 0.0 & 1.4_{-0.2}^{+1.5} \pm 0.0 & (22.6,29.9)  &  24.7 & (3.8 , 4.8)  &  4.1 & 0.1 \\ \hline
3/2 & 1 & 4.8_{-0.1}^{+0.2} \pm 0.1 & 5.7_{-0.0}^{+0.3} \pm 0.3 & 1.8_{-0.4}^{+0.8} \pm 0.1 & 2.2_{-0.0}^{+0.8} \pm 0.1 & (28.2,34.7)  &  30.4 & (3.4 , 4.4)  &  6.7 & 0.3 \\ \hline
5/2 & 1 & 6.0_{-0.4}^{+0.2} \pm 0.1 & 5.9_{-0.3}^{+0.1} \pm 0.0 & 6.9_{-3.9}^{+2.6} \pm 0.1 & 11.6_{-4.7}^{+2.7} \pm 0.1 & (37.6,50.0)  &  45.3 & (3.8 , 4.8)  &  1.7 & 0.6
    \\ \hline
    \end{array}
    $$
    \caption{\label{tab:SR-results-udcuC}
        QCD SR results for the masses $m_\pm$ and the decay constants $f_\pm$ given for a $(udc)$-$(\bar cu)$ pentaquark of both parities
        with spin 1/2, 3/2, 5/2 (first column) with type 1 current for each case (second column).
        See the caption of Tab.~\ref{tab:SR-results} for more details.
    }
\end{table*}

The values of the masses and the decay constants can be extracted from the decoupled QCD SRs, Eq.~(\ref{eq:SRexplicit}), through averaging in the Borel window $M^2\in[M^2_-,M^2_+]$
\begin{eqnarray}\label{eq:massSR}
  m_\pm(s_0) &=&\frac{1}{n+1}\sum\limits_{j=0}^n m_\pm(s_0,M^2_j)\,,\\\nn
f^2_\pm(s_0) &=&\frac{1}{n+1}\sum\limits_{j=0}^n
e^{m_\pm^2/M^2_j}{\cal R}^\text{(SR)}_{\pm,0}(M^2_j,s_0)\,,
\end{eqnarray}
where $n=8$, $M^2_j=M^2_{-}+(M^2_{+}-M^2_-)j/n$ and
\begin{eqnarray}\label{eq:massSRborel}
m^{2\Delta k}_\pm(s_0,M^2) &=&
\frac{{\cal R}^\text{(SR)}_{\pm,k+\Delta k}(M^2,s_0)}
{{\cal R}^\text{(SR)}_{\pm,k         }(M^2,s_0)}
\,.
\end{eqnarray}
We present our result for the case $(k,\Delta k)=(1/2,1/2)$.
We have also checked two extra choices:
$(0,1/2)$ and $(1/2,1)$ for $(k,\Delta k)$ to
confirm the small dependence of our results on $k$ and $\Delta k$.
Similar decoupled QCD SRs have been considered in~\cite{Wang:2015epa} with $(k,\Delta k)=(1/2,1)$.
Borel parameter dependencies of the masses $m_\pm(s_0,M^2)$
for the $uds-{\bar c}c$ case are shown in Fig. \ref{fig:fig-SR}
for the best-fit threshold value $s_0=\tilde s_0$.
Additionally, the bands around the central value show the dependence of the masses on the threshold $s_0$
varied in the interval $s_0\in[s_0^\text{min},s_0^\text{max}]$.

To find the best values of the five parameters $f_\pm$, $m_\pm$, $s_0$,
we demand the minimization of the Borel parameter dependence of the original coupled SRs, Eqs.~(\ref{eq:SR12})
 i.e., 
\begin{eqnarray}\nn
\delta_i^k(s_0)=
\!\!\!\!\mathop{\textbf{max}}\limits_{M^2\in[M^2_-,M^2_+]}
\frac{
    {\cal R}^\text{(res)}_{i,k}(M^2)-{\cal R}^\text{(SR)}_{i,k}(M^2,s_0)}
{{\cal R}^\text{(res)}_{i,k}(M^2)}
\cdot 100\%
\end{eqnarray}
with masses and decay constants in ${\cal R}^\text{(res)}$  fixed by Eqs.~(\ref{eq:massSR}).
The minimization of the Borel parameter dependence of the original coupled SRs instead of the decoupled SRs
helps avoiding possible uncertainties
related to the analytical continuation of the SRs.
Finally, we combine the four criteria in one to get
\begin{eqnarray}\nn
\delta(s_0)=\textbf{max}(
\delta_1^k(s_0),\delta_1^{k+\Delta k}(s_0),
\delta_{2}^{k-\frac 12}(s_0),\delta_{2}^{k+\Delta k-\frac 12}(s_0))\,.
\end{eqnarray}
We use this combined criterion to define the best-fit value for the threshold $\tilde{s_0}$
and the threshold interval $s_0\in[s_0^\text{min},s_0^\text{max}]$,
where subject to the condition
\begin{eqnarray}\nn
\delta(s_0)<\delta(\tilde s_0)+1\,.
\end{eqnarray}
The values of the threshold $\tilde s_0$ and the
interval boundaries $s_0^\text{min}$ and $s_0^\text{max}$
can be found in Tables~\ref{tab:SR-results}, \ref{tab:SR-results-udcsC}, \ref{tab:SR-results-udcuC}
for all considered states.
From these values we obtain the masses $m_\pm(\tilde s_0)$ and the decay constant  $f_\pm(\tilde s_0)$
at $\tilde s_0$ given in Table~\ref{tab:SR-results}
together with their variations in the threshold interval
and the variations in the Borel window.

The central value $\bar m$ of the mass and the
uncertainty $\Delta_{s} m$ related to the threshold
are defined by
\begin{eqnarray}\nn
\bar m_\pm &=&\frac 12
\left(
    \mathop{\textbf{max}}\limits_{s_0}~m_\pm(s_0)+
    \mathop{\textbf{min}}\limits_{s_0}~m_\pm(s_0)
\right)\\\nn
\Delta_{s} m_\pm &=&\frac 12
\left(
\mathop{\textbf{max}}\limits_{s_0}~m_\pm(s_0)-
\mathop{\textbf{min}}\limits_{s_0}~m_\pm(s_0)
\right)\,,
\end{eqnarray}
where $\textbf{max}$ ($\textbf{min}$) gives the maximum (minimum) value of the function $m_\pm(s_0)$
in the threshold interval $s_0\in[s_0^\text{min},s_0^\text{max}]$.
The error bars related to the Borel parameter
variation in the Borel window interval $M^2\in[M^2_-,M^2_+]$ is calculated by
\begin{eqnarray}\nn
\Delta_{M} m_\pm &=& \frac 12
\left(
\mathop{\textbf{max}}\limits_{M^2}~m_\pm(\tilde s_0,M^2)-
\mathop{\textbf{max}}\limits_{M^2}~m_\pm(\tilde s_0,M^2)
\right)\,.
\end{eqnarray}
Final results for the mass are given in Fig. \ref{fig:fig-Allmasses} and Tab.~\ref{tab:SR-results-short} by the central value mass $\bar m$
and the total uncertainty $\Delta m$
\begin{eqnarray}\label{eq:massSRcentral}
m_\pm=\bar m_\pm+\Delta m_\pm\,,
\end{eqnarray}
where the total uncertainty is the sum of the above uncertainties
\begin{eqnarray}\label{eq:massSRerror}
\Delta m_\pm &=&\Delta_{s} m_\pm + \Delta_{M} m_\pm
\end{eqnarray}
that includes only uncertainties stemming from the SR analysis
and don't include the uncertainties of the condensates.

The following numerical values of the vacuum condensates
and masses have been used for the numerical analysis
\begin{eqnarray}\nn
&&\va{(\al_S/\pi)G^2}= 0.012~\text{GeV}^3\,,~~
\va{\bar qq}= (-0.25)^3~\text{GeV}^3\,,~~\\\nn
&&\va{\bar qGq}=\va{\bar q G_{\mu\nu}\sigma_{\mu\nu}q}=m_0^2\va{\bar qq},
~
m_0^2=0.8~\text{GeV}^2\,,~~\\\nn
&& m_q =0\,,~~m_s = 0.1~\text{GeV}\,,~~m_c = 1.23~\text{GeV}\,,~~ ~~\\\nn
&&\va{\bar ss} = f_s \va{\bar qq}\,,~~\va{\bar sGs}=f_s \va{\bar qGq}\,,~~
f_s =0.8\,.
\end{eqnarray}
The lowest threshold value is taken to be $s_\text{th}=6.5$~GeV$^2$,
see Eq. (\ref{eq:SR-disp}).

The QCD SR technique described above has been applied using various pentaquark currents.
First, we studied the type-1 current for the three flavor configuration ($uds$-$\bar cc$, $udc$-$\bar cs$, $usc$-$\bar cd$), see the results in Table~\ref{tab:SR-results}.
Second, in Table~\ref{tab:SR-results-udcsC}, we obtained results for some alternative currents to estimate their relevance.
Finally, we used our method to study the $udc$-$\bar cu$
flavor configuration, in order to see whether $P_c^+(4312)$, $P_c^+(4440)$, and $P_c^+(4457)$,
observed by the LHCb Collaboration, can be understood as a pentaquark of two clusters in a color-octet state.
The detailed results given in Table~\ref{tab:SR-results-udcuC} will be discussed in the next section.

\section{Discussion and Summary}
In this section, we discuss the results obtained in the previous sections on the basis of the constructed QCD SRs for pentaquark states.
We have constructed the currents for $udsc\bar c$ pentaquarks of spin-1/2, 3/2, 5/2 that have two clusters of a color-octet.
The first cluster consists of three quarks $q^1q^2q^3$ which has the same flavor structure as the flavor singlet state of $uds$,
while the second cluster consists of quark-antiquark $\bar cq^4$.
There are four options for flavor clustering
$q^1q^2q^3$-$\bar cq^4$ ($uds$-$\bar cc$, $udc$-$\bar cs$, $usc$-$\bar cd$, $dsc$-$\bar cu$).
The results for $dsc$-$\bar cu$ and $usc$-$\bar cd$ are identical in our approach and, therefore, we present here only results for the $dsc$-$\bar cu$ configuration.
The main predictions for pentaquarks are presented for the type-1 current,
that has a spin-1 $\bar cq^4$ part.
In section \ref{sec:OPE}, in addition to these main currents,
we have also introduced the alternative currents
for spin-1/2 states and spin-3/2 states,
see Eq.~(\ref{eq:current12types}) and
Eq.~(\ref{eq:current32types}).
Particularly, we are interested in the alternative currents with a spin-0 quark-antiquark cluster:
type-3 and type-4 for a spin-1/2 current and type-2 for a spin-3/2 current.
In Table~\ref{tab:SR-results-udcsC}, we presented the results for these alternative currents of a $udc$-$\bar cs$ configuration
with a spin-0 $\bar cs$-cluster in
comparison with the main currents that have a spin-1
$\bar cs$-cluster.
One can see that these types of currents lead to larger masses
compared to those for the spin-1 cases for both spin-1/2 and spin-3/2 pentaquarks.
We have also checked that a similar conclusion is valid for other flavor configurations.
This observation agrees with~\cite{Irie:2017qai}, where it has been shown that the two-quark cluster with spin 1 in $uds$-${\bar c}c$ system yield the most stable result.
In Table~\ref{tab:SR-results-udcsC}, we have also considered the alternative current for a spin 1/2 state containing a spin-1 $\bar cs$-cluster (type-2 for spin-1/2 current)
and found that this current gives the same result.
Therefore, the main results in our paper are given for the hidden pentaquark states with a spin-1 quark-antiquark cluster.

\begin{figure}[t]
    \includegraphics[width=0.48\textwidth]{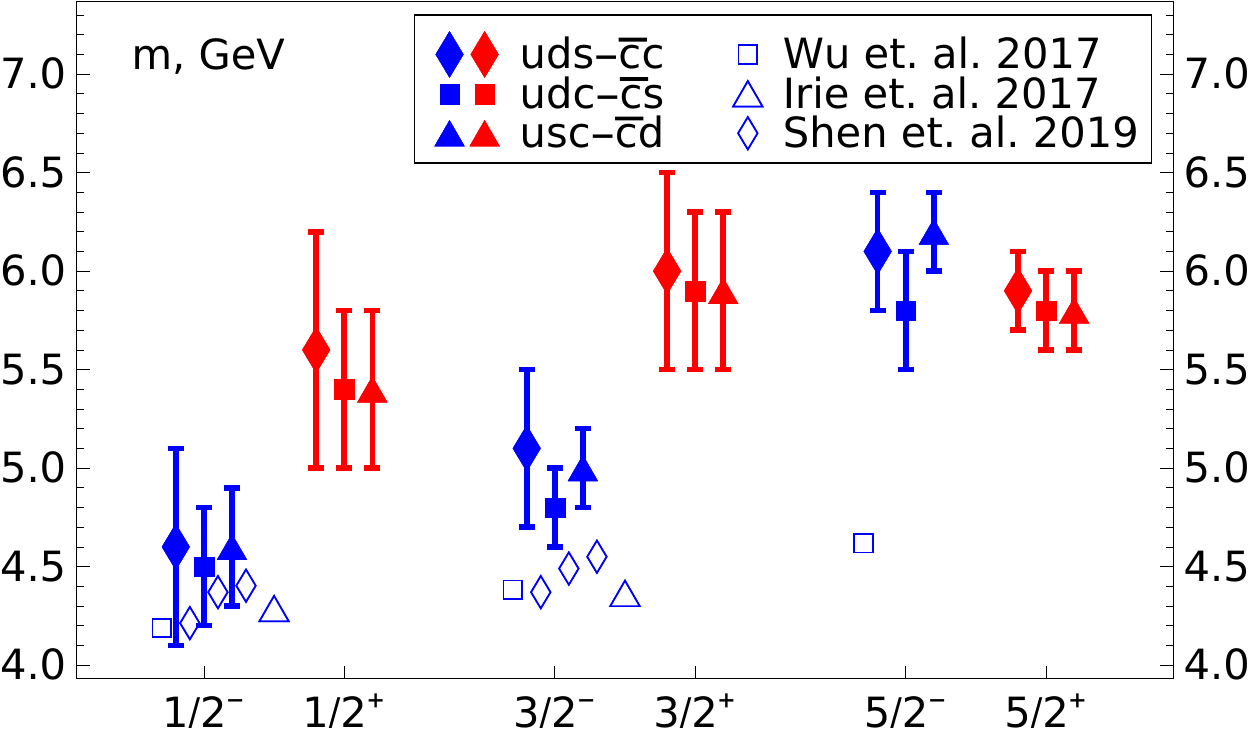}
    \caption{
        \label{fig:fig-Allmasses}
        QCD SRs results for masses of pentaquarks
        with spins 1/2, 3/2, 5/2
        for the even parity (red color errorbars) and
        for the odd parity (blue color errorbars)
        are given for three types of flavor clustering:
        $uds$-$\bar cc$ (diamonds),
        $udc$-$\bar cs$ (squares),
        $usc$-$\bar cd$ (triangles).
        Central value and width of errorbars are given in Eq.~(\ref{eq:massSRcentral})
        and Eq.~(\ref{eq:massSRerror}).
        The result of our calculations are depicted
        by $\blacklozenge$ for $uds$-$\bar cc$,
        by $\blacksquare$  for $udc$-$\bar cs$,
        by $\blacktriangle$ for $usc$-$\bar cd$.
        The results of other theoretical predictions for $udsc\bar c$ pentaquark
        are denoted by  $\square$ for the color-magnetic interaction based study~\cite{Wu:2017weo},
        $\lozenge$ for the framework of the coupled channel unitary approach with the local hidden gauge formalism~\cite{Wu:2010jy,Wu:2010vk,Shen:2019evi},
        $\triangle$ for the quark model result~\cite{Irie:2017qai}.
        }
\end{figure}

\begin{figure}[b]
    \includegraphics[width=0.48\textwidth]{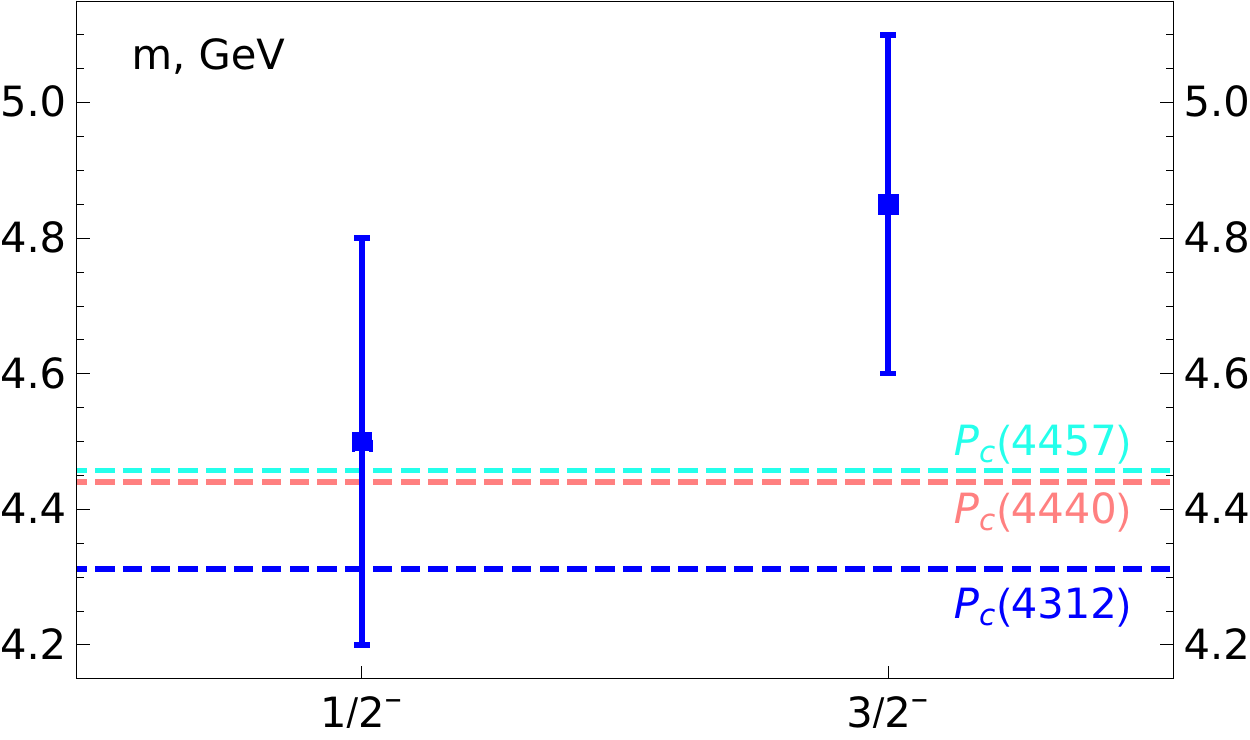}
    \caption{
        \label{fig:fig-Allmasses-LHCb}
        The masses of a recently observed by LHCb~\cite{Aaij:2019vzc} states are shown by the dashed lines in comparison to our QCD SR estimations (blue errorbars) for the lightest states with a color-octet substructure.
            For more details see Tab.~\ref{tab:SR-results-udcuC}.
    }
\end{figure}

Using type-1 currents, we have considered three types of flavor clustering
($uds$-$\bar cc$, $udc$-$\bar cs$, $usc$-$\bar cd$) and
found that they have similar masses and decay constants, see Table~\ref{tab:SR-results}.
Therefore, we expect that these configurations have equal chances to be observed.
The consideration of a possible mixing between these configurations is outside the scope of this work.
Another observation is that the larger spin states give larger masses.

\begin{table}[t]
    \centering
    $$
    \begin{array}{|c||c|c||c|c||c|c|} \hline
    \text{flavor}
    & 1/2^- & 1/2^+ & 3/2^- & 3/2^+ & 5/2^- & 5/2^+\\ \hline\hline
uds\text{-}\bar cc  &   4.6(5) & 5.6(6)  &  5.1(4) & 6.0(5)  &  6.1(3) & 5.9(2) \\ \hline
udc\text{-}\bar cs  &   4.5(3) & 5.4(4)  &  4.8(2) & 5.9(4)  &  5.8(3) & 5.8(2) \\ \hline
usc\text{-}\bar cd  &   4.6(3) & 5.4(4)  &  5.0(2) & 5.9(4)  &  6.2(2) & 5.8(2) \\ \hline
    \end{array}
    $$
    \caption{\label{tab:SR-results-short}
        Final QCD SR results for $udsc\bar c$ pentaquark masses
        for both parities with spin 1/2, 3/2, 5/2.
        Values are given according Eq.~(\ref{eq:massSRcentral}).
        For more details see Table~\ref{tab:SR-results}.
}
\end{table}

Our results are presented in comparison with
other theoretical predictions~\cite{Shen:2019evi,Wu:2017weo,Irie:2017qai}
for $uds\bar cc$ pentaquark
in Fig. \ref{fig:fig-Allmasses}.
The masses from the effective Lagrangian framework~\cite{Shen:2019evi}
for the $uds$-$\bar cc$ flavor configuration
with a color-singlet substructures,
depicted by~$\lozenge$, are lower for the spin 3/2 case and 
comparable consistently well with our predictions for the spin 1/2 case 
referring to a pentaquark state with a color-octet substructure.
The quark model prediction for $uds$-${\bar c}c$~\cite{Irie:2017qai}, noted by $\triangle$ in
Fig.~\ref{fig:fig-Allmasses},
is in very good agreement with our result for a spin-1/2 pentaquark, 
while the  prediction for spin-3/2 case is different.
Note that apart of result~\cite{Irie:2017qai}, we compare our predictions with the results
for the configurations that are different from the configurations considered in our work.
Therefore, the comparisons are given only for the reference.

In order to see whether
any of the pentaquarks observed by the LHCb Collaboration could be understood as a pentaquark composed of two clusters in the color-octet state,
we study the pentaquark with the assumption that it is formed by the two clusters $udc$-$\bar cu$, where the three-quark cluster
has a flavor-singlet structure.
Therefore, we don't have alternative to $udc$-$\bar cu$
flavor clustering as opposite to the $uds\bar cc$ pentaquark.
QCD SR results for the masses $m_\pm$ and decay constants $f_\pm$ for such a pentaquark are
presented in Table~\ref{tab:SR-results-udcuC} for spin 1/2, 3/2, 5/2 and both parities.
To make a point, we present the lightest state masses from this table in Fig.~\ref{fig:fig-Allmasses-LHCb} together with the states recently observed by the LHCb Collaboration.
As shown in this figure, the obtained mass for a spin-1/2 $udc$-$\bar cu$ pentaquark, is in agreement with the experimental value. 
Since the flavor content and the mass of the pentaquarks observed by LHCb are known only,
we conclude that
if the observed state has spin 1/2 and negative parity, it could be described as a state with two color-octet clusters.

	It had been shown~\cite{Kondo:2004cr,Lucha:2019pmp} that 
	the correlator for a pentaquark state could include the two-hadron-reducible contributions, which are given by convolution of baryon and meson correlators that is not related to pentaquark.
	This problem of QCD SRs has been addressed in the series of work~\cite{Lucha:2019pmp,Lucha:2019cpe,Lucha:2019cdc}
	for tetraquark QCD SRs, where authors expect that similar problem could affects also
	SR for pentaquarks.
	As it has been shown for pentaquark considered in~\cite{Lee:2004xk,Agaev:2019qqn}, the direct subtraction of a problematic two-hadron contributions from the correlator leads to incorrect results.
	To avoid this problem, the authors of~\cite{Lee:2004xk} utilized soft-kaon theorem and
	demonstrated that these type of problematic terms contribute less than 10\% of the sum rules.
We propose that the type of pentaquark currents constructed in this work
is a solution for this problem of pentaquark SRs,
due to the fact that such currents can not be factorized to the product of meson and baryon currents, see the relevant discussion in Appendix C.




To summarize, we have estimated the masses of the
various hidden-charm pentaquarks with color-octet substructure and with $J^{PC}=$1/2$^\pm$, 3/2$^\pm$, 5/2$^\pm$
in the framework of QCD SRs.
We have constructed the currents for a particular configuration of pentaquark states,
which consists of a three-quark cluster with the same flavor structure as the flavor singlet combination $uds$,
and, additionally, of a quark-antiquark cluster, where both clusters are in a color-octet state.
In our work, three possible types of flavor-clustering of the currents has been considered.
To obtain QCD sum rules, the operator product expansion for the correlators with the constructed interpolating
currents has been performed up to the level of dimension-10 condensates.
From the constructed QCD SRs the masses and decay constants of the pentaquark states
have been extracted.
Numerical values are given in detail in Table~\ref{tab:SR-results},
and are briefly summarized in Table~\ref{tab:SR-results-short}.

\section{Acknowledgment}

We would like to thank N. Stefanis, M. Elbistan, D. Melikhov, Ju-Jun Xie and Hua-Xing Chen for stimulating discussions and useful remarks.
We were inspired to perform this study by Nikolai Kochelev, who passed suddenly away in 2018.
This work was supported by the National Natural Science Foundation of China (Grant No. 11975320),
the National Key Research and Development Program of China (No. 2016YFE0130800),
the Chinese Academy of Sciences President's International Fellowship Initiative
(PIFI Grant No. 2019PM0036).
The work of H.-J. Lee was supported by the Basic Science Research Program through the National Research
Foundation of Korea (NRF) funded by
the Ministry of Education under Grant No. 2016R1D1A1A09920078.
The work has been partially supported by the Ministry of Education and Science of the Russian Federation: Projects No. 3.6371.2017/8.9 and No. 3.6439.2017/8.9.

\begin{appendix}

\section{Projectors for 3/2-spin correlator}
\label{app:32projectors}
We follow the common practice to extract the $g_{\mu\nu}$-term of the $3/2$ correlator and consider only the largest spin contribution. Here, we formalize this extraction by introducing the appropriate 
projectors.
The general form of the tensor can be written in the following way
\begin{eqnarray}\label{eq:pi32expansion}
P_{\mu\nu}(q^2)&=& \sum\limits_{i=1}^{5}(c_{1i} \hat q + c_{2i})t^i_{\mu\nu}=
\sum\limits_{j=1}^{10}C_j \tilde t^j_{\mu\nu}\,,
\end{eqnarray}
where we consider only P-even terms.
The relation between two forms is given by
$\tilde t^{2i}=t^i$, $\tilde t^{2i-1}=\hat qt^i$,
$C_{2i}=c_{2i}$, $C_{2i-1}=c_{1i}$ with $i=1,\cdots\,5$.
The linearly independent set $t^j_{\mu\nu}$ of all possible structures is defined as follows
\begin{eqnarray}\nn
t^j_{\mu\nu}=\left(-g_{\mu\nu},\ga_\mu\ga_\nu,\frac{q_\mu q_\nu}{q^2},q_\mu\ga_\nu-q_\nu\ga_\mu,q_\mu\ga_\nu+q_\nu\ga_\mu\right)_j\,.
\end{eqnarray}
A linear combination of tensors $\tilde t_{\mu\nu}$ can be used to construct the projectors as
\begin{eqnarray}\nn
P^{3/2,k}_{\mu\nu}=M^{-1}_{kl}\tilde t^l_{\mu\nu}\,,
\end{eqnarray}
where the matrix $M_{kl}$ reads
\begin{eqnarray}\nn
M_{kl}=Tr\left(\tilde t^l_{\mu\nu}\tilde t^k_{\mu\nu}\right)\,.
\end{eqnarray}
Then we can extract the coefficients $C_j$ from the expansion expressed by Eq.~\ref{eq:pi32expansion}
\begin{eqnarray}\nn
C_k=Tr\left(P_{\mu\nu}(q^2)P^{3/2,k}_{\mu\nu}\right).
\end{eqnarray}
The inverse of the matrix $M$ is given by
\begin{eqnarray}\nn
&&24s^2 M^{-1}=\\\nn
&&\small
\left(
\begin{array}{cccccccccc}
s & 0 & -s & 0 & 2 s & 0 & 0 & s & 0 & 0 \\
0 & s^2 & 0 & -s^2 & 0 & 2 s^2 & s & 0 & 0 & 0 \\
-s & 0 & -s & 0 & 0 & 0 & 0 & s & 0 & 0 \\
0 & -s^2 & 0 & -s^2 & 0 & 0 & s & 0 & 0 & 0 \\
2 s & 0 & 0 & 0 & 12 s & 0 & 0 & 0 & 0 & -2 s \\
0 & 2 s^2 & 0 & 0 & 0 & 4 s^2 & 0 & 0 & 2 s & 0 \\
0 & s & 0 & s & 0 & 0 & -2 & 0 & 0 & 0 \\
s & 0 & s & 0 & 0 & 0 & 0 & 0 & 0 & 0 \\
0 & 0 & 0 & 0 & 0 & 2 s & 0 & 0 & -1 & 0 \\
0 & 0 & 0 & 0 & -2 s & 0 & 0 & 0 & 0 & s \\
\end{array}
\right)\,,
\end{eqnarray}
where $s=q^2$.
Using the projectors $P^{3/2,1}_{\mu\nu}$ and $P^{3/2,2}_{\mu\nu}$,
the densities, Eq.~(\ref{eq:rho-32}), can recast in the form
\begin{eqnarray}\nn
\rho^{3/2}_1(s)&=&
\frac{-1}{24\pi s}{\rm Tr}\bigg[\textbf{Im}\Pi^{3/2}_{\mu\nu}(s)
\hat{q}\bigg(
  g^{\mu\nu}+\ga^\mu\ga^\nu-\frac{2q^\mu q^\nu}{q^2}
  \bigg)\bigg]
\\&&\nn
+\frac{1}{24\pi s}{\rm Tr}\bigg(\textbf{Im}\Pi^{3/2}_{\mu\nu}(s)(q^\mu\gamma^\nu-q^\nu\gamma^\mu)\bigg)\,,
\\\nn
\rho^{3/2}_2(s)
&=&\frac{-1}{24\pi}{\rm Tr}\bigg[\textbf{Im}\Pi^{3/2}_{\mu\nu}(s)
\bigg(
  g^{\mu\nu}+\ga^\mu\ga^\nu-\frac{2q^\mu q^\nu}{q^2}
\bigg)\bigg]
\\&&    \label{eq:rho-32-mod}
-\frac{1}{24\pi s}{\rm Tr}\bigg(\textbf{Im}\Pi^{3/2}_{\mu\nu}(s)\hat{q}(q^\mu\gamma^\nu-q^\nu\gamma^\mu)\bigg)\,.
\end{eqnarray}

\section{Projectors for 5/2-spin correlator}
\label{app:52projectors}
To extract the terms of the largest spin state from the correlator, the projector method is applied.
Similarly to Eq.~(\ref{eq:pi32expansion}), the general form of the correlator can be written as follows
\begin{eqnarray}\label{eq:pi52expansion}
P_{\mu\nu,\al\be}(q^2)&=& \sum\limits_{i=1}^{14}(c_{1i} \hat q + c_{2i})t^i_{\mu\nu,\al\be}=
\sum\limits_{j=1}^{28}C_j \tilde t^j\,,
\end{eqnarray}
where the relation between the two forms is given by
$\tilde t^{2i}=t^i_{\mu\nu,\al\be}$, $\tilde t^{2i-1}=\hat qt^i_{\mu\nu,\al\be}$,
$C_{2i}=c_{2i}$, $C_{2i-1}=c_{1i}$ with $i=1,\cdots\,14$.
We consider only P-even terms which are symmetric with respect to $\mu\nu$ and $\al\be$.
The linearly independent set $t^i_{\mu\nu,\al\be}$ of all possible structures is defined as
\begin{eqnarray}\nn
&& \!\!\!\!\!\!
t^i_{\mu\nu,\al\be}=\hat S_{\mu\nu}\hat S_{\al\be}\left(\right.
\frac{g_{\mu\al}g_{\nu\be}}{4},  
\frac{g_{\mu\nu}g_{\al\be}}{4},  
\frac{g_{\mu\al}\ga_\nu\ga_\be}{4},  
g_{\mu\al}\frac{q_\nu q_\be}{q^2},  
\\\nn &&  \!\!
g_{\mu\nu}\frac{q_\al q_\be}{4q^2},  
g_{\al\be}\frac{q_\mu q_\nu}{4q^2},  
g_{\mu\nu}\frac{\ga_\al q_\be}{2},  
g_{\al\be}\frac{\ga_\mu q_\nu}{2},  
g_{\mu\al}q_\nu \ga_\be,  
\\\nn && \!\! \left.
g_{\mu\al}q_\be \ga_\nu,  
\frac{q_\mu q_\nu}{2q^2}\ga_\al q_\be,  
\frac{q_\al q_\be}{2q^2}\ga_\mu q_\nu,  
\frac{q_\mu q_\al}{q^2}\ga_\nu \ga_\be,  
\frac{q_\mu q_\nu q_\al q_\be}{q^4}
\right)_i\,,
\end{eqnarray}
where the operator $\hat  S_{\mu\nu}$ symmetrizes the tensor as ${\hat S_{\mu\nu} t_{\mu\nu} = t_{\mu\nu}+t_{\nu\mu}}$.
Linear combination of these tensors can be used as the projectors
\begin{eqnarray}\label{eq:52projector}
P^{5/2,k}_{\mu\nu,\al\be}=M^{-1}_{kl}\tilde t^l
\end{eqnarray}
to extract the coefficients $C_j$ of the expansion, Eq.~(\ref{eq:pi52expansion}) notably,
\begin{eqnarray}\nn
C_j=Tr\left(P_{\mu\nu,\al\be}P^{5/2,j}_{\mu\nu\al\be}\right)
\end{eqnarray}
with the matrix
\begin{eqnarray}\nn
M_{kl}=Tr\left(\tilde t^l_{\mu\nu,\al\be}\tilde t^k_{\mu\nu,\al\be}\right)\,.
\end{eqnarray}
We provide only the first two rows of the inverse matrix, which define the projectors $P^{5/2,1}$ and $P^{5/2,2}$ applied
to extract the spin-5/2 spectral densities, Eq.~(\ref{eq:rho-52}):
\begin{eqnarray}\nn
M^{-1}_{1l} 120q^2 &=&
(2, 0, -2, 0, 1, 0, -2, 0, 2, 0, 2, 0, 0, 0, 0, 0, 0,
\\\nn&& -1, 0, 1, 0, 2, 0, -2, -1, 0, 4, 0)_l\,,
\\\nn
M^{-1}_{2l} 120q^2 &=& (0, 2q^2, 0, -2q^2, 0, q^2, 0, -2q^2, 0, 2q^2, 0, 2q^2,
\\\nn&& \!\!\!\!\!\!\!\!
0, 0, 0, 0, -1, 0, 1, 0, 2, 0, -2, 0, 0, -q^2, 0, 4q^2)_l\,.
\end{eqnarray}
Other rows of the inverse matrix are not used in our work but could be obtained from the above equations.

\section{Currents in color subspace}

Here, we consider the relation of the pentaquarks with different configurations:
 diquark-diquark-antiquark clustering $J_{\bar 3}\sim qq$-$qq$-$\bar q$
 with an anti-triplet color substructure suggested in \cite{Jaffe:2003sg,Sugiyama:2003zk,Lee:2005ny},
a molecule form $J_1\sim qqq$-$\bar qq$
with color-singlet parts, see~\cite{Guo:2017jvc},
and the combination $J_8\sim qqq$-$\bar qq$ with color-octet compounds studied here.
First, we consider only the color part of these currents
\begin{eqnarray}\nn
J_8^\text{c}&\equiv&  (\epsilon_{a_1a_2a_0}t^m_{a_0a_3}q_1q_2q_3)\cdot(\bar q_5 t^m_{a_5a_4}q_4)\,,\\\nn
J_1^\text{c}&\equiv&  3(\epsilon_{a_1a_2a_4}q_1q_2q_4)\cdot(\bar q_5 \delta_{a_5a_3}q_3)\,,\\\nn
J_{\bar 3}^\text{c}&\equiv&  6(\epsilon_{ia_1a_2}q_1q_2)\cdot(\epsilon_{ja_1a_2}q_3q_4)\cdot \bar q_5
\epsilon_{ija_5}  \,.
\end{eqnarray}
Using a Fiertz identity, one can get the relation
\begin{eqnarray}\nn
J_8^\text{c} =  J_1^\text{c} +  J_{\bar 3}^\text{c}\,,
\end{eqnarray}
where the quark fields carry flavor $f_i$, color $c_i$ and spin $l_i$ indices
as $q_i=q_{f_ic_il_i}$.
Then, multiplying this relation with the same spinor tensor
\begin{eqnarray}
T_{l_1l_2l_3l_4l_5l}=(\Gamma_1)_{l_1l_2}(\Gamma_2)_{ll_3}(\Gamma_3)_{l_5l_4}\,,
\end{eqnarray}
one can obtain a relation between the full currents
\begin{eqnarray}\nn
J_8 =  J_1 +  J_{\bar 3}\,,
\end{eqnarray}
where $J_t =  J_t^\text{c} T_{l_1l_2l_3l_4l_5l}$.
The tensor has been introduced in such a way so that the definition for $J_8$
agrees with Sec.~\ref{sec:interpolating.J}:
\begin{eqnarray}\nn
J_8 =
    \epsilon_{a_1a_2a_0}t^m_{a_0a_3}(q_1\Gamma_1 q_2)
    (\Gamma_2 q_3)_l
    (\bar q_5 t^m \Gamma_3q_4)\,.
\end{eqnarray}
After performing a Fiertz transformation in the currents,
we get:
\begin{eqnarray}\nn
J_1 &=&
3\sum_{N=1}^{5} \epsilon_{a_1a_2a_4}(q_1\Gamma_1 q_2) (\Gamma_2^N q_4)_l
    \bar q_5 \Gamma_3^N q_3\,,\\\nn
J_{\bar 3}&=& 6\sum_{N=1}^{5}
    \epsilon_{ia_1a_2}(q_1\Gamma_1 q_2)
    \epsilon_{ja_1a_2}(q_3\tilde\Gamma_3^N q_4)
    \epsilon_{ija_5}(\tilde\Gamma_2^N q_4)_l  \,,
\end{eqnarray}
where the modified matrices $\tilde\Gamma_i^N$ and $\Gamma_i^N$
are defined in terms of the Fiertz identity
\begin{eqnarray}\nn
\delta_{ij}\delta_{kl}=\sum_{N=1}^{5} \Delta_{il}^N\Delta_{kj}^N\,,
\end{eqnarray}
where
$
\Delta^N=
    ( 1/2\,,~\ga_5/2\,,~ \ga_\rho/2\,,~ i\ga_5\ga_\rho/2\,,~
      i\si_{\rho,\ga}/\sqrt{8}
    )_N\,
$.
Then, the definition for the modified matrices $\tilde\Gamma_i^N$ and $\Gamma_i^N$ take the form
\begin{eqnarray}\nn
\Gamma_i^N= \Gamma_i\Delta^N\,,~~
\tilde\Gamma_2^N=-\Gamma_2\Delta^NC\,,~~
\tilde\Gamma_2^N=(\Delta^N)^TC\Gamma_3\,.
\end{eqnarray}
Therefore, currents with color-octet parts $J_8$, color-singlet parts $J_1$,
and color-anti-triplet parts $J_{\bar 3}$ are linearly dependent.
Including the flavor symmetry into consideration will cause the break of the clustering of $J_1$ and $J_{\bar 3}$. 
In other words, for $J_8$ current with the same factorization in color, spin and flavor, 
the factorization of the currents $J_1$ and $J_{\bar 3}$ in flavor space is differ from the factorization in spin space. 
That could protect the current $J_8$ from being presented as a product of meson and barion currents.
According to our knowledge, the currents of type $J_1$ and $J_{\bar 3}$ with different clustering in flavor and spin 
have not been considered in the literature.
Therefore, we would like to point out
that the currents suggested in our work cannot be a linear combination of
any other currents considered previously,
for example in \cite{Jaffe:2003sg,Sugiyama:2003zk,Lee:2005ny, Guo:2017jvc}.

    \section{Spectral densities}
Here we collect the analytical results for the
spectral densities $\rho^{s}_{D,i}(t,\al,\be)$,
where $s$ denotes spin,
$D$ - the dimension of OPE term,
$i$ - the part of the correlator ($i=1\,,2$).
Here, we present only the result
for the $uds$-$\bar cc$ flavor configuration.
We use notations, $L=t\al\be -m_c^2(\al+\be)$, $\ga=1-\al-\be$,
and $\be_0=(\be_++\be_-)/2$.
The latter notation has been introduced to combine various terms under a two dimensional integral, Eq.~(\ref{eq:rho2Dintegral}), so that
\begin{eqnarray}\nn
\int_{\be_-}^{\be_+}\!\!\!\!d\be\,\de(\be-\be_0)=1\,.
\end{eqnarray}

\begin{widetext}\footnotesize
    \begin{eqnarray}\label{eq:rho-result-12}
\rho^{1/2}_{1,0}(t,\al,\be)&=&(\ga ^3 L^4 (5 \ga  m_c^2+8 L))/(15 \pi ^8 \al ^4 \be ^4 2^{14})\,, ~~
\rho^{1/2}_{2,0}(t,\al,\be)=-(\ga ^2 L^4 m_s (10 \ga  m_c^2+3 L))/(5 \pi ^8 \al ^4 \be ^4 2^{13} 3^{2})\,,
\\ \nn
\rho^{1/2}_{1,3}(t,\al,\be)&=&(\ga  L^2 \va{\bar qq} (3 f_s-2) m_s (3 \ga  m_c^2+4 L))/(\pi ^6 \al ^2 \be ^2 2^{9} 3^{2})
\,, ~~
\rho^{1/2}_{2,3}(t,\al,\be)=(\ga  L^3 \va{\bar qq} (f_s+2) (4 \ga  m_c^2+L))/(\pi ^6 \al ^3 \be ^3 2^{10} 3^{2})\,,
\\ \nn
\rho^{1/2}_{1,4}(t,\al,\be)&=&(\va{(\al_s/\pi) GG} \ga  L (\ga  L m_c^2 (-32 \al ^3 \ga +3 \al ^2 (12 \be ^2+7 \be  \ga +4 \ga ^2)+6 \al  \be ^2 \ga +4 \be ^2 \ga  (3 \ga -8 \be ))-8 \ga ^3 (\al ^3+\be ^3) m_c^4
\\ \nn  &&
+3 \al  \be  L^2 (16 \al  \be +7 \al  \ga +2 \be  \ga )))/(\pi ^6 \al ^4 \be ^4 2^{15} 3^{2})\,,
\\ \nn
\rho^{1/2}_{2,4}(t,\al,\be)&=&-(\va{(\al_s/\pi) GG} L m_s (6 \ga  L m_c^2 (-8 \al ^3 \ga +\al ^2 (30 \be ^2+21 \be  \ga +16 \ga ^2)+6 \al  \be ^2 \ga -8 \be ^2 \ga  (\be -2 \ga ))-64 \ga ^3 (\al ^3+\be ^3) m_c^4
\\ \nn  &&
+3 \al  \be  L^2 (10 \al  \be +28 \al  \ga +8 \be  \ga +\ga ^2)))/(\pi ^6 \al ^4 \be ^4 2^{16} 3^{3})\,,
\\ \nn
\rho^{1/2}_{1,5}(t,\al,\be)&=&-(L \va{\bar qGq} m_s (\ga  m_c^2+L) (-8 \al  \be +7 \al  \ga +2 \be  \ga +16 \al  \be  f_s))/(3 \pi ^6 \al ^2 \be ^2 2^{12})\,,
\\ \nn
\rho^{1/2}_{2,5}(t,\al,\be)&=&(5 \ga  L^3 \va{\bar qGq} (\al -\be ) (f_s+2))/(\pi ^6 \al ^3 \be ^3 2^{13} 3^{2})\,,
~
\rho^{1/2}_{1,6}(t,\al,\be) = (L \va{\bar qq}^2 (2 f_s+1) (\ga  m_c^2+L))/(\pi ^4 \al  \be  2^{5} 3^{2})\,,
\\ \nn
\rho^{1/2}_{2,6}(t,\al,\be)&=&(L \va{\bar qq}^2 m_c^2 (f_s-6) m_s)/(\pi ^4 \al  \be  2^{4} 3^{2})-(\va{\bar qq}^2 \delta(\be -\be_0) (f_s-6) m_s (m_c^2+(\al -1) \al  t)^2)/(\pi ^4 (\al -1) \al  2^{6} 3^{2})\,,
\\ \nn
\rho^{1/2}_{1,8}(t,\al,\be)&=&(\va{\bar qGq} \va{\bar qq} (2 f_s+1) (m_c^2 (-4 \al  \be +7 \al  \ga +2 \be  \ga )+L (7 \al +2 \be )))/(\pi ^4 \al  \be  2^{9} 3^{2})
\\ \nn  &&
+(\va{\bar qGq} \va{\bar qq} \delta(\be -\be_0) (2 f_s+1) (m_c^2+(\al -1) \al  t))/(\pi ^4 2^{6} 3^{2})\,,
\\ \nn
\rho^{1/2}_{2,8}(t,\al,\be)&=&(\va{\bar qGq} \va{\bar qq} \delta(\be -\be_0) m_s (m_c^2 (5 (2 \al -1) f_s-4 (5 \al +2))
+(\al -1) \al  t ((2 \al  (14 \al -9)-5) f_s
\\ \nn  &&
-4 (\al  (72 \al -67)+2))))/(\pi ^4 (\al -1) \al  2^{10} 3^{2})
-(\va{\bar qGq} \va{\bar qq} (7 \al +2 \be ) m_c^2 m_s)/(\pi ^4 \al  \be  2^{7} 3^{2})\,,
\\ \nn
\rho^{1/2}_{1,9}(t,\al,\be)&=&0\,,~~~~ 
\rho^{1/2}_{2,9}(t,\al,\be)=-((\al -1) \al  \va{\bar qq}^3 t \delta(\be -\be_0) f_s)/(3 \pi ^2 2^{2})\,,
\\ \nn
\rho^{1/2}_{1,10}(t,\al,\be)&=&-((\al -1) \al  \va{\bar qGq}^2 \delta(\be -\be_0) (2 f_s+1))/(3 \pi ^4 2^{8})\,, ~
\rho^{1/2}_{2,10}(t,\al,\be)=((\al -1) \al  \va{\bar qGq}^2 \delta(\be -\be_0) m_s)/(\pi ^4 2^{7}) \,,
\\
\label{eq:rho-result-32}
\rho^{3/2}_{1,0}(t,\al,\be)&=&(\ga ^3 L^4 (5 (5-2 \ga ) \ga  m_c^2+(13 \ga +4) L))/(5 \pi ^8 \al ^4 \be ^4 2^{15} 3^{2})\,,
\\ \nn
\rho^{3/2}_{2,0}(t,\al,\be)&=&-(\ga ^2 L^4 m_s (5 (5-2 \ga ) \ga  m_c^2+(11 \ga +3) L))/(5 \pi ^8 \al ^4 \be ^4 2^{14} 3^{3})\,,
\\ \nn
\rho^{3/2}_{1,3}(t,\al,\be)&=&(\ga  L^2 \va{\bar qq} (3 f_s-2) m_s (3 (5-2 \ga ) \ga  m_c^2+(9 \ga +2) L))/(\pi ^6 \al ^2 \be ^2 2^{10} 3^{3})\,,
\\ \nn
\rho^{3/2}_{2,3}(t,\al,\be)&=&(\ga  L^3 \va{\bar qq} (f_s+2) (4 (5-2 \ga ) \ga  m_c^2+(9 \ga +2) L))/(\pi ^6 \al ^3 \be ^3 2^{12} 3^{3})\,,
\\ \nn
\rho^{3/2}_{1,4}(t,\al,\be)&=&-(\va{(\al_s/\pi) GG} \ga  L (6 \ga  L m_c^2 (8 \al ^3 (\ga  (7 \ga +4)-18 \be ^2)+\al ^2 (-72 (2 \be +3) \be ^2+(\be -72) \ga ^2+42 \be  \ga )+\al  \be ^2 \ga  (\ga +12)
\\ \nn  &&
+8 \be ^2 \ga  (7 \be  \ga +4 \be -9 \ga ))+96 (\ga -1) \ga ^3 (2 \ga -5) m_c^4 ((\ga -1)^2-3 \al  \be )+\al  \be  L^2 (\ga  (72 \be +\ga  (9 \ga -20))
\\ \nn  &&
-36 \al  (36 \be  \ga +8 \be -7 \ga ))))/(\pi ^6 \al ^4 \be ^4 2^{18} 3^{4})\,,
\\ \nn
\rho^{3/2}_{2,4}(t,\al,\be)&=&(\va{(\al_s/\pi) GG} L m_s (6 \ga  L m_c^2 (\al ^3 (-90 \be ^2+42 \be  \ga +8 \ga  (5 \ga +3))+3 \al ^2 (-15 (2 \be +3) \be ^2+12 (\be -2) \ga ^2+(18 \be +7) \be  \ga )
\\ \nn  &&
+\al  \be ^2 \ga  (12 \be +11 \ga +6)+8 \be ^2 \ga  (\be  (5 \ga +3)-9 \ga ))-24 (\ga -1) \ga ^3 (8 \al ^3+\al ^2 (7 \be +12)+2 \al  (\be -6) \be +4 \be ^2 (2 \be +3)) m_c^4
\\ \nn  &&
+\al  \be  L^2 (-90 \al  \be  (7 \ga +1)+42 \al  (7 \ga +2) \ga +12 \be  (7 \ga +2) \ga +(7 \ga -15) \ga ^2)))/(\pi ^6 \al ^4 \be ^4 2^{17} 3^{5})\,,
\\ \nn
\rho^{3/2}_{1,5}(t,\al,\be)&=&(L \va{\bar qGq} m_s (\ga  m_c^2 (8 \al  \be  (5-2 \ga )+7 \al  \ga +2 \be  \ga +16 \al  \be  (2 \ga -5) f_s)+L (4 \al  \be  (7 \ga +1)+7 \al  \ga +2 \be  \ga
\\ \nn  &&
-8 \al  \be  (7 \ga +1) f_s)))/(\pi ^6 \al ^2 \be ^2 2^{13} 3^{2})\,,
\\ \nn
\rho^{3/2}_{2,5}(t,\al,\be)&=&-(5 \ga  L \va{\bar qGq} (\al -\be ) (f_s+2) (\ga  (4 \ga +3) L m_c^2-8 (\ga -1) \ga ^2 m_c^4+(7 \ga +2) L^2))/(\pi ^6 \al ^3 \be ^3 2^{15} 3^{4})\,,
\\ \nn
\rho^{3/2}_{1,6}(t,\al,\be)&=&(L \va{\bar qq}^2 (2 f_s+1) (2 (5-2 \ga ) \ga  m_c^2+7 \ga  L+L))/(\pi ^4 \al  \be  2^{7} 3^{3})\,,
\\ \nn
\rho^{3/2}_{2,6}(t,\al,\be)&=&(L \va{\bar qq}^2 (f_s-6) m_s ((10-4 \ga ) m_c^2+5 L))/(\pi ^4 \al  \be  2^{7} 3^{3})-(\va{\bar qq}^2 \delta(\be -\be_0) (f_s-6) m_s (m_c^2+(\al -1) \al  t)^2)/(\pi ^4 (\al -1) \al  2^{7} 3^{3})\,,
\\ \nn
\rho^{3/2}_{1,8}(t,\al,\be)&=&(\va{\bar qGq} \va{\bar qq} \delta(\be -\be_0) (2 f_s+1) (m_c^2+(\al -1) \al  t))/(\pi ^4 2^{8} 3^{3})-(\va{\bar qGq} \va{\bar qq} (2 f_s+1) (m_c^2 (\al  \be  (20-8 \ga )+7 \al  \ga +2 \be  \ga )
\\ \nn  &&
+L (\al  (20 \be +7)+2 \be )))/(\pi ^4 \al  \be  2^{10} 3^{3})\,,
\\ \nn
\rho^{3/2}_{2,8}(t,\al,\be)&=&(\va{\bar qGq} \va{\bar qq} m_s (4 (7 \al +2 \be ) ((5-2 \ga ) m_c^2+3 L)-5 (\al -\be ) f_s (m_c^2+3 L)))/(\pi ^4 \al  \be  2^{11} 3^{4})
+(\va{\bar qGq} \va{\bar qq} \delta(\be -\be_0) m_s (m_c^2 (20 \al
\\ \nn  &&
+(5-10 \al ) f_s+8)+(\al -1) \al  t (4 (437-432 \al ) \al +(2 \al  (84 \al -89)+5) f_s+8)))/(\pi ^4 (\al -1) \al  2^{11} 3^{4})\,,
\\ \nn
\rho^{3/2}_{1,9}(t,\al,\be)&=&0\,, ~~~ 
\rho^{3/2}_{2,9}(t,\al,\be)=-((\al -1) \al  \va{\bar qq}^3 t \delta(\be -\be_0) f_s)/(\pi ^2 2^{2} 3^{2})\,,
\\ \nn
\rho^{3/2}_{1,10}(t,\al,\be)&=&-((\al -1) \al  \va{\bar qGq}^2 \delta(\be -\be_0) (2 f_s+1))/(\pi ^4 2^{9} 3^{2})\,,
~ 
\rho^{3/2}_{2,10}(t,\al,\be)=((\al -1) \al  \va{\bar qGq}^2 \delta(\be -\be_0) m_s)/(\pi ^4 2^{9})\,,
    \end{eqnarray}
    \begin{eqnarray}\label{eq:rho-result-52}
\rho^{5/2}_{1,0}(t,\al,\be)&=&(\ga ^3 L^3 (\ga  ((139-104 \ga ) \ga +45) L m_c^2+8 (\ga -1) \ga ^2 (4 \ga -9) m_c^4+2 (13 \ga  (3 \ga +1)+6) L^2))/(3 \pi ^8 \al ^4 \be ^4 2^{13} 5^{2})\,,
\\ \nn
\rho^{5/2}_{2,0}(t,\al,\be)&=&(\ga ^2 L^3 m_s (5 \ga  m_c^2 (8 (\ga -1) \ga  (4 \ga -9) m_c^2+((117-88 \ga ) \ga +36) L)+2 (11 \ga  (13 \ga +4)+18) L^2))/(\pi ^8 \al ^4 \be ^4 2^{13} 3^{2} 5^{2})\,,
\\ \nn
\rho^{5/2}_{1,3}(t,\al,\be)&=&(\ga  L \va{\bar qq} m_s (3 \ga  L m_c^2 (-40 \ga +((95-72 \ga ) \ga +27) f_s+100)+12 (\ga -1) \ga ^2 (4 \ga -9) m_c^4 f_s
\\ \nn  &&
+2 L^2 (90 \ga +9 (\ga  (11 \ga +3)+1) f_s+20)))/(5 \pi ^6 \al ^2 \be ^2 2^{8} 3^{2})\,,
\\ \nn
\rho^{5/2}_{2,3}(t,\al,\be)&=&(\ga  L^2 \va{\bar qq} (f_s+2) (2 \ga  m_c^2 ((\ga  (72 \ga -95)-27) L-6 (\ga -1) \ga  (4 \ga -9) m_c^2)-9 (\ga  (11 \ga +3)+1) L^2))/(5 \pi ^6 \al ^3 \be ^3 2^{8} 3^{3})\,,
\\ \nn
\rho^{5/2}_{1,4}(t,\al,\be)&=&(\va{(\al_s/\pi) GG} \ga  (-3 \ga  L^2 m_c^2 (8 \al ^3 (180 \be ^2 (3-7 \ga )+225 \be  \ga +4 \ga  (7 \ga  (9 \ga +5)+30))+\al ^2 (-1440 \be ^3 (7 \ga -3)
+100 \be ^2 (69-13 \ga )
\\ \nn  &&
+\be  \ga  (\ga  (72 \ga +1495)+900)-240 \ga ^2 (7 \ga +5))+\al  \be ^2 \ga  (1800 \be +\ga  (72 \ga +1495)+900)
+16 \be ^2 \ga  (2 \be  (7 \ga  (9 \ga +5)+30)
\\ \nn  &&
-15 \ga  (7 \ga +5)))+12 (\ga -1) \ga ^2 L m_c^4 (32 \al ^3 \ga  (14 \ga +5)
+\al ^2 (140 \be ^2 (4 \ga -9)+\be  \ga  (4 \ga +225)
-40 (\ga -5) \ga )
\\ \nn  &&
+\al  \be  \ga  (\be  (4 \ga +225)-200 (\ga +1))+8 \be ^2 \ga  (56 \be  \ga +20 \be -5 \ga +25))+192 (\ga -1)^2 \ga ^4 (4 \ga -9) m_c^6 ((\ga -1)^2
-3 \al  \be )
\\ \nn  &&
+2 \al  \be  L^3 (30 \al  (\be  (6 \ga  (77 \ga -54)-58)-45 \ga  (3 \ga +1))+\ga  (\ga  (9 (20-11 \ga ) \ga +70)-1350 \be  (3 \ga +1)))))/(\pi ^6 \al ^4 \be ^4 2^{16} 3^{3} 5^{2})\,,
\\ \nn
    \rho^{5/2}_{2,4}(t,\al,\be)&=&
(\va{(\al_s/\pi) GG} m_s (6 \ga  m_c^2 (4 (\ga -1) \ga  m_c^2 (4 (\ga -1) \ga ^2 (4 \ga -9) m_c^2 ((\ga -1)^2-3 \al  \be )+L (16 \al ^3 \ga  (5 \ga +2)
+\al ^2 (21 \be ^2 (4 \ga -9)
\\ \nn  &&
+\be  (\ga -10) \ga +10 (4-3 \ga ) \ga )+\al  \be  \ga  ((\be -30) \ga +40 (\be -1))+2 \be ^2 \ga  (8 \be  (5 \ga +2)-15 \ga +20)))+L^2 (4 \al ^3 (42 \be ^2 (7 \ga -4)
\\ \nn  &&
+15 \be  \ga -2 \ga  (5 \ga  (7 \ga +4)+18))+\al ^2 (168 \be ^3 (7 \ga -4)+15 \be ^2 (23 \ga -70)+2 \be  \ga  ((33-7 \ga ) \ga +15)+60 \ga ^2 (5 \ga +4))
\\ \nn  &&
-2 \al  \be ^2 \ga  (120 \be +\ga  (7 \ga +92)+60)-4 \be ^2 \ga  (\be  (10 \ga  (7 \ga +4)+36)-15 \ga  (5 \ga +4))))+\al  \be  L^3 (588 \al  \be  (\ga -1) (9 \ga +1)
\\ \nn  &&
+60 \al  (7 \ga +2) \ga -240 \be  (7 \ga +2) \ga +7 ((16-9 \ga ) \ga +6) \ga ^2)))/(5 \pi ^6 \al ^4 \be ^4 2^{15} 3^{4})\,,
\\ \nn
    \rho^{5/2}_{1,5}(t,\al,\be)&=&
-(\va{\bar qGq} m_s (24 \al  \be  f_s (\ga  ((73-56 \ga ) \ga +18) L m_c^2+2 (\ga -1) \ga ^2 (4 \ga -9) m_c^4+(7 \ga  (9 \ga +2)+3) L^2)
\\ \nn  &&
+5 L (2 \ga  (2 \ga -5) m_c^2 (\al  (\ga -44 \be )-4 \be  \ga )+L (44 \al  \be  (7 \ga +1)-\al  \ga  (7 \ga +2)+4 \be  \ga  (7 \ga +2)))))/(5 \pi ^6 \al ^2 \be ^2 2^{11} 3^{2})\,,
\\ \nn
    \rho^{5/2}_{2,5}(t,\al,\be)&=&
(L \va{\bar qGq} (f_s+2) (\ga  L m_c^2 (3 \ga  (4 \al ^2 (280 \be +11)+\al  (4 \be  (280 \be +97)+55)+\be  (4 \be +5))-56 \ga ^3 (11 \al +\be )+81 \ga ^2 (11 \al +\be )
\\ \nn  &&
+1080 \al  \be )+4 (\ga -1) \ga ^2 (4 \ga -9) m_c^4 (60 \al  \be +11 \al  \ga +\be  \ga )+2 L^2 (\al  (20 \be  (7 \ga  (9 \ga +2)+3)+11 \ga  (7 \ga  (3 \ga +1)+3))
\\ \nn  &&
+\be  \ga  (7 \ga  (3 \ga +1)+3))))/(5 \pi ^6 \al ^3 \be ^3 2^{12} 3^{3})\,,
\\ \nn
\rho^{5/2}_{1,6}(t,\al,\be)&=&-(L \va{\bar qq}^2 (2 f_s+1) (2 (5-2 \ga ) \ga  m_c^2+7 \ga  L+L))/(\pi ^4 \al  \be  2^{4} 3^{2})\,,
\\ \nn
    \rho^{5/2}_{2,6}(t,\al,\be)&=&
(\va{\bar qq}^2 \delta(\be -\be_0) (f_s+5) m_s (m_c^2+(\al -1) \al  t)^2)/(15 \pi ^4 (\al -1) \al  2^{4})-(\va{\bar qq}^2 m_s (2 L m_c^2 (-30 \ga +((51-40 \ga ) \ga +9) f_s+75)
\\ \nn  &&
+4 (\ga -1) \ga  (4 \ga -9) m_c^4 f_s+5 L^2 (2 (7 \ga +1) f_s+15)))/(5 \pi ^4 \al  \be  2^{4} 3^{2})\,,
\\ \nn
    \rho^{5/2}_{1,8}(t,\al,\be)&=&
(\va{\bar qGq} \va{\bar qq} (2 f_s+1) ((2 \ga -5) m_c^2 (\al  (\ga -22 \be )-4 \be  \ga )+L (\al  (110 \be -5 \ga -1)+4 (5 \be  \ga +\be ))))/(\pi ^4 \al  \be  2^{7} 3^{3})
\\ \nn  &&
-(11 \va{\bar qGq} \va{\bar qq} \delta(\be -\be_0) (2 f_s+1) (m_c^2+(\al -1) \al  t))/(\pi ^4 2^{6} 3^{3})\,,
\\ \nn
    \rho^{5/2}_{2,8}(t,\al,\be)&=&
(\va{\bar qGq} \va{\bar qq} m_s (f_s (m_c^2 (4 \al ^2 (228 \be +66 \ga +11)+2 \al  \be  (456 \be +144 \ga +119)+55 \al  (\ga +1)+\be  (24 \be  \ga +4 \be +5 \ga +5))
\\ \nn  &&
+6 L (\al  (190 \be +55 \ga +11)+5 \be  \ga +\be ))+90 (\ga -1) ((2 \ga -5) m_c^2-3 L)))/(5 \pi ^4 \al  \be  2^{8} 3^{3})
\\ \nn  &&
-(\va{\bar qGq} \va{\bar qq} \delta(\be -\be_0) m_s (m_c^2 ((38 \al ^2-48 \al -1) f_s-30)+(\al -1) \al  t (30 (48 (\al -1) \al -1)
\\ \nn  &&
+(10 \al  (19 \al -20)-1) f_s)))/(5 \pi ^4 (\al -1) \al  2^{8} 3^{2})\,,
\\ \nn
\rho^{5/2}_{1,9}(t,\al,\be)&=&0\,, ~~~ 
\rho^{5/2}_{2,9}(t,\al,\be)=-((\al -1) \al  \va{\bar qq}^3 t \delta(\be -\be_0) f_s)/(3 \pi ^2)\,,
\\ \nn
\rho^{5/2}_{1,10}(t,\al,\be)&=&((\al -1) \al  \va{\bar qGq}^2 \delta(\be -\be_0) (2 f_s+1))/(3 \pi ^4 2^{6})\,,
~~~
\rho^{5/2}_{2,10}(t,\al,\be)=(3 (\al -1) \al  \va{\bar qGq}^2 \delta(\be -\be_0) m_s)/(\pi ^4 2^{7})\,.
    \end{eqnarray}

\end{widetext}

\end{appendix}


\begin{thebibliography}{10}
	
	\bibitem{Aaij:2015tga}
	R. Aaij {\it et~al.}, Phys. Rev. Lett.
	\textbf{115},  072001  (2015). 
	
	\bibitem{Aaij:2019vzc}
	R. Aaij {\it et~al.}, Phys. Rev. Lett.
	\textbf{122},  222001  (2019). 

\bibitem{Cao:2019kst}
X.~Cao and J.~P.~Dai,
Phys.\ Rev.\ D {\bf 100}, no. 5, 054033 (2019)

\bibitem{Cao:2019gqo}
X.~Cao, F.~K.~Guo, Y.~T.~Liang, J.~J.~Wu, J.~J.~Xie, Y.~P.~Xie, Z.~Yang and B.~S.~Zou,
arXiv:1912.12054 

\bibitem{Wang:2019krd}
X.~Y.~Wang, X.~R.~Chen and J.~He,
Phys.\ Rev.\ D {\bf 99}, no. 11, 114007 (2019)
	
	\bibitem{Wu:2017weo}
	J. Wu {\it et~al.}, Phys. Rev.
	\textbf{D95},  034002  (2017). 
	
	\bibitem{Santopinto:2016pkp}
	E. Santopinto and A. Giachino, Phys. Rev.
	\textbf{D96},  014014  (2017). 
	
	\bibitem{Irie:2017qai}
	Y. Irie, M. Oka, and S. Yasui, Phys. Rev.
	\textbf{D97},  034006  (2018). 
	
	\bibitem{Maiani:2015vwa}
	L. Maiani, A.~D. Polosa, and V. Riquer, Phys. Lett.
	\textbf{B749},  289  (2015). 
	
	\bibitem{Lebed:2015tna}
	R.~F. Lebed, Phys. Lett.
	\textbf{B749},  454  (2015). 
	
	\bibitem{Li:2015gta}
	G.-N. Li, X.-G. He, and M. He, JHEP
	\textbf{12},  128  (2015). 
	
	\bibitem{Zhu:2015bba}
	R. Zhu and C.-F. Qiao, Phys. Lett.
	\textbf{B756},  259  (2016). 
	
	\bibitem{Wang:2015epa}
	Z.-G. Wang, Eur. Phys. J.
	\textbf{C76},  70  (2016). 
	
	\bibitem{Jaffe:2003sg}
	R.~L. Jaffe and F. Wilczek, Phys. Rev. Lett.
	\textbf{91},  232003  (2003). 
	
	\bibitem{Sugiyama:2003zk}
	J. Sugiyama, T. Doi, and M. Oka, Phys. Lett.
	\textbf{B581},  167  (2004). 
	
	\bibitem{Lee:2005ny}
	H.-J. Lee, N.~I. Kochelev, and V. Vento, Phys. Rev.
	\textbf{D73},  014010  (2006). 
	
	\bibitem{Karliner:2003dt}
	M. Karliner and H.~J. Lipkin, Phys. Lett.
	\textbf{B575},  249  (2003). 
	
	\bibitem{Zhu:2003ba}
	S.-L. Zhu, Phys. Rev. Lett.
	\textbf{91},  232002  (2003). 
	
	\bibitem{Chen:2015moa}
	H.-X. Chen {\it et~al.}, Phys. Rev. Lett.
	\textbf{115},  172001  (2015). 
	
	\bibitem{Roca:2015dva}
	L. Roca, J. Nieves, and E. Oset, Phys. Rev.
	\textbf{D92},  094003  (2015). 
	
	\bibitem{He:2015cea}
	J. He, Phys. Lett.
	\textbf{B753},  547  (2016). 
	
	\bibitem{Azizi:2016dhy}
	K. Azizi, Y. Sarac, and H. Sundu, Phys. Rev.
	\textbf{D95},  094016  (2017). 
	
	\bibitem{Guo:2017jvc}
	F.-K. Guo {\it et~al.}, Rev. Mod. Phys.
	\textbf{90},  015004  (2018). 
	
	\bibitem{Wu:2010jy}
	J.-J. Wu, R. Molina, E. Oset, and B.~S. Zou, Phys. Rev. Lett.
	\textbf{105},  232001  (2010). 
	
	\bibitem{Wu:2010vk}
	J.-J. Wu, R. Molina, E. Oset, and B.~S. Zou, Phys. Rev.
	\textbf{C84},  015202  (2011). 
	
	\bibitem{Shen:2019evi}
	C.-W. Shen, J.-J. Wu, and B.-S. Zou, arXiv:1906.03896
	(2019). 
	
	\bibitem{Liu:2019tjn}
	M.-Z. Liu {\it et~al.}, Phys. Rev. Lett.
	\textbf{122},  242001  (2019). 
	
	\bibitem{Eides:2017xnt}
	M.~I. Eides, V.~{\relax Yu}. Petrov, and M.~V. Polyakov, Eur. Phys. J.
	\textbf{C78},  36  (2018). 
	
	\bibitem{Chen:2016qju}
	H.-X. Chen, W. Chen, X. Liu, and S.-L. Zhu, Phys. Rept.
	\textbf{639},  1  (2016). 
	
	\bibitem{Mironov:2015ica}
	A. Mironov and A. Morozov, JETP Lett.
	\textbf{102},  271  (2015). 
	Teor. Fiz.102,no.5,302(2015)].
	
	\bibitem{Takeuchi:2016ejt}
	S. Takeuchi and M. Takizawa, Phys. Lett.
	\textbf{B764},  254  (2017). 
	
	\bibitem{Ioffe:1981kw}
	B.~L. Ioffe, Nucl. Phys.
	\textbf{B188},  317  (1981). 
	Phys.B191,591(1981)].
	
	\bibitem{Ioffe:2010zz}
	B.~L. Ioffe, V.~S. Fadin, and L.~N. Lipatov, {\em {Quantum chromodynamics:
			Perturbative and nonperturbative aspects}}
	(Cambridge Univ. Press, ADDRESS, 2010), Vol.~30. 
	
	\bibitem{Gursey:1992dc}
	F. Gursey and L.~A. Radicati, Phys. Rev. Lett.
	\textbf{13},  173  (1964). 
	
	\bibitem{bookCloseQuarks}
	F.~E. Close, An introduction to quarks and partons, (Academic press, 1982)
	
	
	\bibitem{Shifman:1978bx}
	M.~A. Shifman, A.~I. Vainshtein, and V.~I. Zakharov, Nucl. Phys.
	\textbf{B147},  385  (1979). 
	
	\bibitem{Chung:1981cc}
	Y. Chung, H.~G. Dosch, M. Kremer, and D. Schall, Nucl. Phys.
	\textbf{B197},  55  (1982). 
	
	\bibitem{Jido:1996ia}
	D. Jido, N. Kodama, and M. Oka, Phys. Rev.
	\textbf{D54},  4532  (1996). 
	
	\bibitem{Leinweber:1989hh}
	D.~B. Leinweber, Annals Phys.
	\textbf{198},  203  (1990). 
	
	\bibitem{Albuquerque:2013ija}
	
	R.~M. Albuquerque, Ph.D. thesis, Sao Paulo U., 2013. 
	ARXIV:1306.4671;
	
	\bibitem{Mikhailov:1986be}
	S.~V. Mikhailov and A.~V. Radyushkin, JETP Lett.
	\textbf{43},  712  (1986). 
	Fiz.43,551(1986)].
	
	\bibitem{Mikhailov:1991pt}
	S.~V. Mikhailov and A.~V. Radyushkin, Phys. Rev.
	\textbf{D45},  1754  (1992). 
	
	\bibitem{Grozin:1985wj}
	A.~G. Grozin and {\relax Yu}.~F. Pinelis, Phys. Lett.
	\textbf{166B},  429  (1986). 
	
	\bibitem{Grozin:1994hd}
	A.~G. Grozin, Int. J. Mod. Phys.
	\textbf{A10},  3497  (1995). 
	
	\bibitem{Bakulev:2006wz}
	A.~P. Bakulev and A.~V. Pimikov, Acta Phys. Polon.
	\textbf{B37},  3627  (2006). 
	
	\bibitem{Bakulev:2009ib}
	A.~P. Bakulev, A.~V. Pimikov, and N.~G. Stefanis, Phys. Rev.
	\textbf{D79},  093010  (2009). 
	
	\bibitem{Xiang:2017byz}
	J.-B. Xiang {\it et~al.}, Chin. Phys.
	\textbf{C43},  034104  (2019). 
	
	\bibitem{Ohtani:2012ps}
	K. Ohtani, P. Gubler, and M. Oka, Phys. Rev.
	\textbf{D87},  034027  (2013). 
	
	\bibitem{Kondo:2004cr}
	Y. Kondo, O. Morimatsu, and T. Nishikawa, Phys. Lett.
	\textbf{B611},  93  (2005). 
	
	\bibitem{Lucha:2019pmp}
	W. Lucha, D. Melikhov, and H. Sazdjian, Phys. Rev.
	\textbf{D100},  014010  (2019). 
	
	\bibitem{Lucha:2019cpe}
	W. Lucha, D. Melikhov, and H. Sazdjian, arxiv:1909.06324
	(2019). 
	
	\bibitem{Lucha:2019cdc}
	W. Lucha, D. Melikhov, and H. Sazdjian, arXiv:1908.10164
	(2019). 
	
	\bibitem{Lee:2004xk}
	S.~H. Lee, H. Kim, and Y. Kwon, Phys. Lett.
	\textbf{B609},  252  (2005). 
	
	\bibitem{Agaev:2019qqn}
	S.~S. Agaev, K. Azizi, and H. Sundu, Phys. Rev.
	\textbf{D99},  114016  (2019). 
	
\end{thebibliography}


\end{document}